\documentclass[12pt]{article}
\usepackage{amssymb}
\usepackage{amsmath}
\usepackage{url}

\usepackage{graphicx}

\usepackage[usenames,dvipsnames]{color}

\begin{document}

\begin{titlepage}
\strut\hfill UMTG--292
\vspace{.5in}
\begin{center}

\LARGE Quantum Integrability and Quantum Groups:\\ 
a special issue in memory of Petr P. Kulish\\
\vspace{1in}
\large 
Nikolai Kitanine \footnote{IMB UMR5584, CNRS, Univ. Bourgogne 
Franche-Comt\'e, F-21000 Dijon, France, Nikolai.Kitanine@u-bourgogne.fr}
, Rafael I. Nepomechie \footnote{Physics Department,
P.O. Box 248046, University of Miami, Coral Gables, FL 33124 USA, 
nepomechie@miami.edu}, 
and Nicolai Reshetikhin \footnote{Department of Mathematics, Evans Hall 917,
University of California, Berkeley, CA 94720-3840 USA, reshetik@math.berkeley.edu}\\[0.8in]
\end{center}

\vspace{.5in}

\begin{abstract}
    This is an introduction to Quantum Integrability and Quantum Groups, a special
issue collection of articles published in Journal of Physics A in memory of 
Petr P. Kulish. A list of Kulish's publications is included.
\end{abstract}

\end{titlepage}

\setcounter{footnote}{0}

\section{Petr P. Kulish: a brief scientific biography}\label{sec:bio}

With deep regret we dedicate the present issue of Journal of Physics A
to the memory of our late colleague and friend Professor Petr Kulish.
Petr Petrovich Kulish was born on February 24, 1944 in Leningrad.  He
passed away on January 14 2016.  Kulish graduated from Leningrad
University (now St.  Petersburg University) in 1966.  He worked all
his life at the Leningrad (later St.  Petersburg) branch of the
Mathematical Institute of the Academy of Science.  He received his PhD
in 1971, and the habilitation degree in 1983.  He became a Professor
of St.  Petersburg University in 1991.  From 2000 he was head of the
Laboratory of Mathematical Methods in Theoretical Physics, which had
been founded and headed by Professor L. D. Faddeev since 1972.

He was one of the first students of L. D. Faddeev.  His first
significant contribution to mathematical physics was a joint paper
with Faddeev on the infrared problem in Quantum Electrodynamics where
they proposed the construction of infrared asymptotic states.  His
research interests ranged from quantum field theory to the theory of
solitons, classical and quantum integrable systems, and algebraic
structures related to integrable systems, such as quantum groups and
related algebras. (A complete list of his publications can be found 
in Section \ref{sec:pubs}.)

During the 70's he was in the center of activity both on classical and
on quantum soliton theory.  His contributions consist of several
important results.  Among them are connections of conservation laws
and asymptotic scattering, semiclassical quantization of the Bose gas
system, and the deformation of conservation laws.  Together with
younger colleagues A. Reiman and V. Gerdzhikov, he developed the
general theory of recursion operators in the classical theory of
solitons.

From the beginning of the 80's Professor Kulish actively worked on the
Quantum Inverse Scattering Method (QISM), algebraic Bethe Ansatz, the
R-matrix approach and on quantum groups.  The very first example of a
quantum group appeared in his joint paper with N. Reshetikhin in 1981,
where the XXZ integrable magnetic chain of higher spin was introduced
in the framework of algebraic Bethe Ansatz.  His early joint reviews
with E. Sklyanin on the R-matrix method and on the Yang-Baxter equation
were very influential at the time and still remain an important
reference.  The fusion method for quantum R-matrices developed in a
joint paper with N. Reshetikhin and E. Sklyanin became one of the
principal tools in the representation theory of quantum groups.  His
other highly important contributions are the multicomponent Bethe
Ansatz, as well as the supersymmetric version of the QISM.

Petr Kulish continued to work very actively on the theory of quantum
groups and its applications to integrable systems in the 90's and in
the 2000's.  Among his results are the studies of the q-oscillators of
the algebraic systems connected with the reflection equations, as well
as of the quantization of the Lorentz group (with eventual
applications to quantization of space-time).

Besides his research, Professor Kulish had outstanding qualities as a
teacher.  Kulish had many graduate students, some of whom became
notable researchers.  He was an invited speaker in numerous
international conferences and schools.  He held visiting positions in
many universities in Finland, France, Portugal, Spain, Sweden and
Italy where he would establish very active collaborations with local
scientists.  Mathematical physics was his life.  He was the second
senior member of the Laboratory after L. D. Faddeev, where he started
to work as a graduate student and finished as its director.  For
all who knew him all these years it is difficult to imagine this place
without Petr.  He is survived by his wife, Olga Kulish, his daughters
Anna and Tatjana, and his grandchildren Maxim and Veronica.  He will be
deeply missed by his friends and colleagues.


\begin{description}
    \item Irina Aref'eva
    \item Eugene Damaskinsky
    \item Nikolai Kitanine
    \item Rafael I. Nepomechie
    \item Nicolai Reshetikhin
    \item Michael Semenov-Tian-Shansky
    \item Alexander Stolin
\end{description}
    
\begin{figure}[h]
\centering
\includegraphics[width=6.0cm]{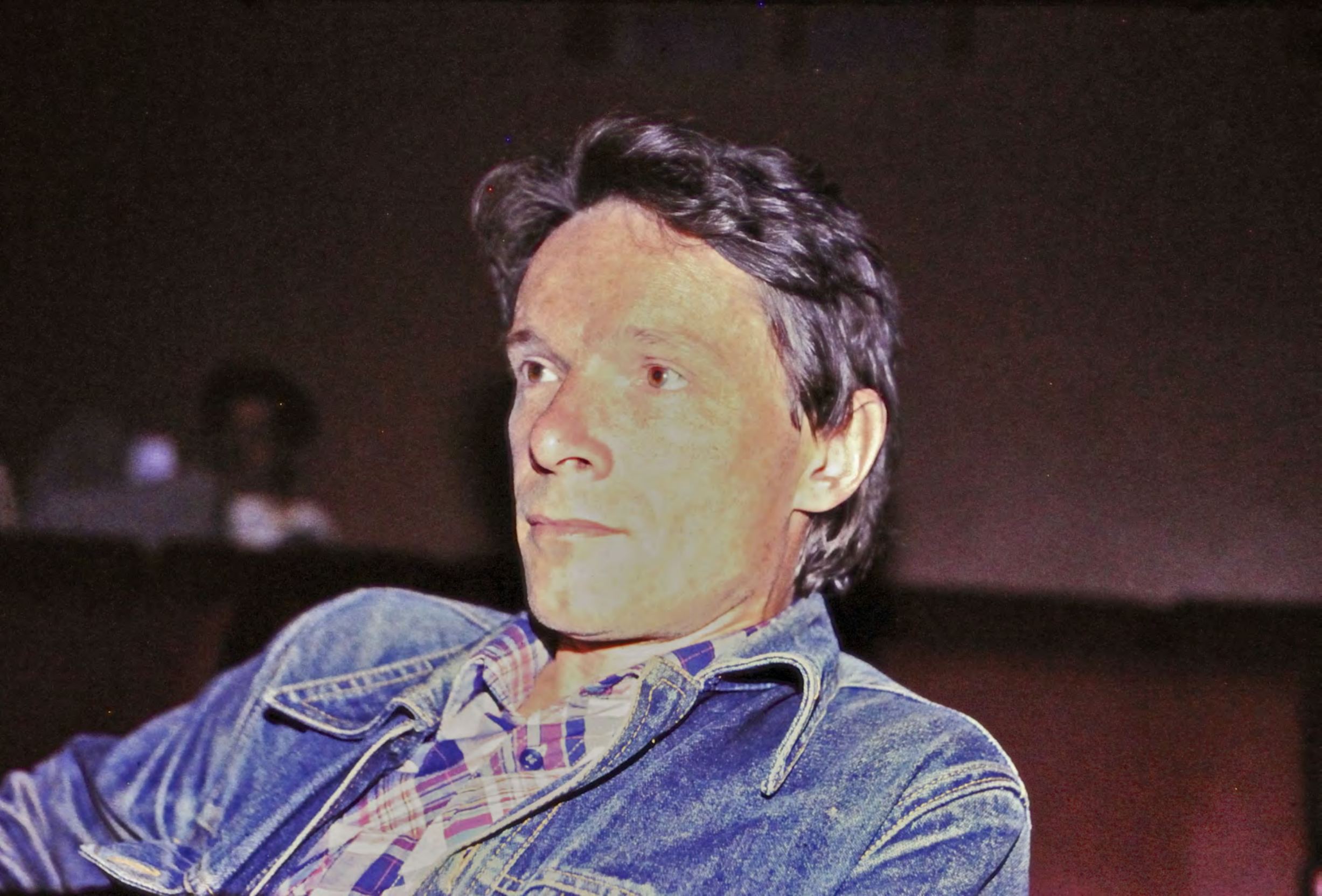}
\caption{Petr Petrovich Kulish. Photo credit: Michael Semenov-Tian-Shansky}
\label{fig:Kulish}
\end{figure}

\section{Contents of the special issue collection}\label{sec:contents}

We are very grateful to the many friends and colleagues of Petr 
Kulish who contributed to this special issue collection. Their 
enthusiastic support for this remembrance, and the breadth 
of subjects covered by their articles, are a testament to the deep 
and wide impact of Kulish and his work. The articles in the collection fall 
into roughly four areas, all of which benefited from seminal 
contributions by Kulish: integrable classical models, 
integrable quantum field theories, integrable quantum lattice models, and 
quantum groups. Brief introductions to these articles are presented 
below.

\subsection{Integrable classical models}

A finite-dimensional {\em superintegrable} model, with non-abelian
integrals of motion, is constructed and investigated in
\cite{Arutyunov:2016kve}.  New techniques for studying soliton
solutions in the Kadomtsev - Petviashvili (KP) II model are developed
in \cite{Boiti:2017}.  The scattering by integrable defects of
finite-gap (quasi-periodic) solutions to the sine-Gordon and Korteweg
- de Vries (KdV) models is studied in \cite{Corrigan:2016sjq}.
B\"acklund-gauge transformations for the KdV hierarchy are proposed in
\cite{Gomes:2016jdg}.  Hamiltonian reduction and other techniques from
Poisson geometry and geometry of Poisson Lie groups are used in
\cite{Feher:2017cic} to derive various multiparticle integrable
systems.  Integrable deformations of sigma models and their
connection to non-abelian duality are investigated in
\cite{Hoare:2016wsk}.

\subsection{Integrable quantum field theories}

The sausage model is an integrable deformation of the two-dimensional
$O(3)$ nonlinear sigma model; and nonlinear integral equations
describing the ground state of this QFT are proposed in
\cite{Ahn:2017ztt}.  Integrable boundary conditions for the
two-dimensional $O(N)$ nonlinear sigma model, both at the classical
and quantum levels, are analyzed in \cite{Aniceto:2017jor}.  Form
factors for the $O(2n)$ Gross-Neveu model are constructed in
\cite{Babujian:2017zoj}.  Form factors for the $SU(N) \times SU(N)$
Principal Chiral Field model are obtained in \cite{Frolov:2017ird}.
In \cite{Bastianello:2017rke}, the non-relativistic limits of the
Gross-Neveu model and the supersymmetric sinh-Gordon and nonlinear
sigma models are investigated.  A certain topological CFT is
explicitly analyzed in \cite{Belavin:2016rav}. A new method of computation of  vacuum expectation values (VEV) in massive integrable field theories is proposed in \cite{Blondeau-Fournier:2016rtu}, it leads, in particular, to some new predictions for the quantum entanglement entropy. Macdonald refined topological vertex introduced and studied in \cite{Foda:2017tnv} is a combinatorial construction closely related to the instanton partition function and AGT conjecture

Several contributions explore integrability in planar AdS/CFT.
Three-point functions in ${\mathcal N}=4$ super Yang-Mills theory (SYM) are computed in 
\cite{Jiang:2016ulr}. Evidence for integrability in the null dipole deformation of 
${\mathcal N}=4$ SYM is found in \cite{Guica:2017mtd}. One-point functions in
a defect CFT are evaluated in \cite{deLeeuw:2016ofj}.

\subsection{Integrable quantum lattice models}

In \cite{Feigin:2017}, XXZ models associated with quantum toroidal
algebras are investigated; and a set of Bethe equations are
conjectured to describe the spectrum of non-local integrals of motion
for the quantum KdV model.  Scalar products of Bethe vectors in models
with $gl(2|1)$ symmetry are computed in \cite{Hutsalyuk:2016ndz,
Hutsalyuk:2016yii}.  A new family of integrable Markov processes,
based on corresponding stochastic R-matrices, has recently been
introduced; and a matrix product formula for the steady-state
probabilities is obtained in \cite{Kuniba:2016iam}.  A fast approach
for determining solutions of the Bethe equations for $gl(n|m)$ spin
chains is presented in \cite{Marboe:2016yyn}.  The $sl(2)$ Hirota
equation is shown to admit a Lax representation with inhomogeneous
terms in \cite{Fioravanti:2016bmi}.  Integrable $A_{2n}^{(2)}$ open
spin chains with quantum group symmetry are studied in
\cite{Ahmed:2017mqq}. A rational R-matrix that takes values in the adjoint representation 
of $su(n)$, and the corresponding integrable spin chain Hamiltonian, are 
constructed in \cite{Stronks:2016nml}.
A novel way of ``closing'' open spin chains based on the 
Temperley-Lieb algebra is studied in \cite{Belletete:2017gwt}.
The construction and solution of integrable models describing boson 
tunneling in multi-well systems are presented in \cite{Ymai:2017yau}. 
An integrable model of quantum nonlinear optics is analyzed in 
\cite{Bogoliubov:2017}. A new Bethe ansatz solution for the relativistic quantum Toda chain is presented in \cite{Zhang:2016kas}.
An unusual type of scale invariance on the lattice is explored in 
\cite{Jones:2017}.
Gustafson integrals and their generalisations frequently appear in the framework of the Separation of Variables for quantum spin chains. Several new results for this type of integrals are obtained in \cite{Derkachov:2016dhc} for the compact case and in \cite{Derkachov:2016ucn} for the non-compact case. The separation of variables method is also used in \cite{Kitanine:2016pvg} to obtain new scalar product  formulas and form factors for open spin chains with non-diagonal boundaries. The study of form factors for the quantum transfer matrix eigen-states is the subject of \cite{Dugave:2016hep}, where this approach is used to study the correlation function in the zero temperature limit. The stability of the spinon excitations for the spin chain with a weak non-integrable perturbation is explored in \cite{Groha:2017wpn}. The spectrum of integrable super-spin chains is considered in \cite{Frahm:2017qfh} for various boundary conditions. The two-dimensional dimer mode is studied in \cite{Pearce:2016vgu} through its correspondence to the free-fermion point of the six-vertex model.

\subsection{Quantum groups}

Distinguished quasi-trigonometric solutions of
the classical Yang-Baxter equation are found in \cite{Burban:2017}.
The $q$-deformation of maximally extended $sl(2|2)$ is investigated by
means of a contraction limit in \cite{Beisert:2017xqx}.  The dynamical
analog of central elements of elliptic quantum algebras is studied in
\cite{Avan:2017xzz}. 
A formula for the scalar product of 
semiclassical eigenvectors of two integrable systems on the same 
symplectic manifold, which leads to the Ponzano-Regge type of 
asymptotic of Racah-Wigner coefficients, is found in \cite{Reshetikhin:2017}.

\section{Petr P. Kulish: publications}\label{sec:pubs}

The list of publications of Petr P. Kulish, compiled using MathSciNet, 
ISI Web of Science and a list produced by Kulish himself in 2004, can be found below 
[42]-[266] \nocite{*}.

\providecommand{\href}[2]{#2}\begingroup\raggedright\endgroup


\begin{thebibliography}{100}

\bibitem{Arutyunov:2016kve}
G.~Arutyunov, M.~Heinze, and D.~Medina-Rincon, ``{Superintegrability of
  Geodesic Motion on the Sausage Model},''
  \href{http://dx.doi.org/10.1088/1751-8121/aa6e0c}{{\em J. Phys.} {\bfseries
  A50} no.~24, (2017) 244002},
\href{http://arxiv.org/abs/1608.06481}{{\ttfamily arXiv:1608.06481 [hep-th]}}.

\bibitem{Boiti:2017}
M.~Boiti, F.~Pempinelli, and A.~K. Pogrebkov, ``{KPII: Cauchy-Jost function,
  Darboux transformations and totally nonnegative matrices},''
  \href{http://dx.doi.org/10.1088/1751-8121/aa7900}{{\em J. Phys.} {\bfseries
  A50} (2017) 304001}, \href{http://arxiv.org/abs/1611.04198}{{\ttfamily
  arXiv:1611.04198 [nlin.SI]}}.

\bibitem{Corrigan:2016sjq}
E.~Corrigan and R.~Parini, ``{Type I integrable defects and finite-gap
  solutions for KdV and sine-Gordon models},''
  \href{http://dx.doi.org/10.1088/1751-8121/aa7612}{{\em J. Phys.} {\bfseries
  A50} no.~28, (2017) 284001},
\href{http://arxiv.org/abs/1612.06904}{{\ttfamily arXiv:1612.06904 [hep-th]}}.

\bibitem{Gomes:2016jdg}
J.~F. Gomes, A.~L. Retore, and A.~H. Zimerman, ``{Miura and generalized
  BŠcklund transformation for KdV hierarchy},''
  \href{http://dx.doi.org/10.1088/1751-8113/49/50/504003}{{\em J. Phys.}
  {\bfseries A49} no.~50, (2016) 504003},
\href{http://arxiv.org/abs/1610.02303}{{\ttfamily arXiv:1610.02303 [nlin.SI]}}.

\bibitem{Feher:2017cic}
L.~Feher and I.~Marshall, ``{The actionÐangle dual of an integrable Hamiltonian
  system of RuijsenaarsÐSchneiderÐvan Diejen type},''
  \href{http://dx.doi.org/10.1088/1751-8121/aa7934}{{\em J. Phys.} {\bfseries
  A50} no.~31, (2017) 314004},
\href{http://arxiv.org/abs/1702.06514}{{\ttfamily arXiv:1702.06514 [math-ph]}}.

\bibitem{Hoare:2016wsk}
B.~Hoare and A.~A. Tseytlin, ``{Homogeneous Yang-Baxter deformations as
  non-abelian duals of the AdS${}_5$ sigma-model},''
  \href{http://dx.doi.org/10.1088/1751-8113/49/49/494001}{{\em J. Phys.}
  {\bfseries A49} no.~49, (2016) 494001},
\href{http://arxiv.org/abs/1609.02550}{{\ttfamily arXiv:1609.02550 [hep-th]}}.

\bibitem{Ahn:2017ztt}
C.~Ahn, J.~Balog, and F.~Ravanini, ``{Nonlinear integral equations for the
  sausage model},'' \href{http://dx.doi.org/10.1088/1751-8121/aa7780}{{\em J.
  Phys.} {\bfseries A50} no.~31, (2017) 314005},
\href{http://arxiv.org/abs/1701.8933}{{\ttfamily arXiv:1701.8933 [hep-th]}}.

\bibitem{Aniceto:2017jor}
I.~Aniceto, Z.~Bajnok, T.~Gombor, M.~Kim, and L.~Palla, ``{On integrable
  boundaries in the 2 dimensional $O(N)$ $\sigma$-models},''
  \href{http://dx.doi.org/10.1088/1751-8121/aa8205}{{\em J. Phys.} {\bfseries
  A50} no.~36, (2017) 364002},
\href{http://arxiv.org/abs/1706.05221}{{\ttfamily arXiv:1706.05221 [hep-th]}}.

\bibitem{Babujian:2017zoj}
H.~M. Babujian, A.~Foerster, and M.~Karowski, ``{Bethe ansatz and exact form
  factors of the O(6) Gross Neveu-model},''
  \href{http://dx.doi.org/10.1088/1751-8121/aa7c7f}{{\em J. Phys.} {\bfseries
  A50} no.~33, (2017) 334003},
\href{http://arxiv.org/abs/1703.05973}{{\ttfamily arXiv:1703.05973 [hep-th]}}.

\bibitem{Frolov:2017ird}
S.~Frolov, ``{Free field representation of the ZF algebra of the $SU(N) \times
  SU(N)$ PCF model},'' \href{http://dx.doi.org/10.1088/1751-8121/aa8226}{{\em
  J. Phys.} {\bfseries A50} no.~37, (2017) 374001},
\href{http://arxiv.org/abs/1705.02602}{{\ttfamily arXiv:1705.02602 [hep-th]}}.

\bibitem{Bastianello:2017rke}
A.~Bastianello, A.~De~Luca, and G.~Mussardo, ``{Non Relativistic Limit of
  Integrable QFT with fermionic excitations},''
  \href{http://dx.doi.org/10.1088/1751-8121/aa6f69}{{\em J. Phys.} {\bfseries
  A50} no.~23, (2017) 234002},
\href{http://arxiv.org/abs/1701.06542}{{\ttfamily arXiv:1701.06542 [hep-th]}}.

\bibitem{Belavin:2016rav}
A.~Belavin and V.~Belavin, ``{On exact solution of topological CFT models based
  on KazamaÐSuzuki cosets},''
  \href{http://dx.doi.org/10.1088/1751-8113/49/41/41LT02}{{\em J. Phys.}
  {\bfseries A49} no.~41, (2016) 41LT02},
\href{http://arxiv.org/abs/1606.05366}{{\ttfamily arXiv:1606.05366 [hep-th]}}.

\bibitem{Blondeau-Fournier:2016rtu}
O.~Blondeau-Fournier and B.~Doyon, ``{Expectation values of twist fields and
  universal entanglement saturation of the free massive boson},''
  \href{http://dx.doi.org/10.1088/1751-8121/aa7492}{{\em J. Phys.} {\bfseries
  A50} no.~27, (2017) 274001},
\href{http://arxiv.org/abs/1612.04238}{{\ttfamily arXiv:1612.04238 [hep-th]}}.

\bibitem{Foda:2017tnv}
O.~Foda and J.-F. Wu, ``{A Macdonald refined topological vertex},''
  \href{http://dx.doi.org/10.1088/1751-8121/aa7605}{{\em J. Phys.} {\bfseries
  A50} no.~29, (2017) 294003},
\href{http://arxiv.org/abs/1701.08541}{{\ttfamily arXiv:1701.08541 [hep-th]}}.

\bibitem{Jiang:2016ulr}
Y.~Jiang, S.~Komatsu, I.~Kostov, and D.~Serban, ``{Clustering and the
  Three-Point Function},''
  \href{http://dx.doi.org/10.1088/1751-8113/49/45/454003}{{\em J. Phys.}
  {\bfseries A49} no.~45, (2016) 454003},
\href{http://arxiv.org/abs/1604.03575}{{\ttfamily arXiv:1604.03575 [hep-th]}}.

\bibitem{Guica:2017mtd}
M.~Guica, F.~Levkovich-Maslyuk, and K.~Zarembo, ``{Integrability in
  dipole-deformed ${\cal N}=4$ super Yang-Mills},''
  \href{http://dx.doi.org/10.1088/1751-8121/aa8491}{{\em J. Phys.} {\bfseries
  A50} no.~39, (2017) 394001},
\href{http://arxiv.org/abs/1706.07957}{{\ttfamily arXiv:1706.07957 [hep-th]}}.

\bibitem{deLeeuw:2016ofj}
M.~de~Leeuw, C.~Kristjansen, and G.~Linardopoulos, ``{One-point functions of
  non-protected operators in the SO(5) symmetric D3ÐD7 dCFT},''
  \href{http://dx.doi.org/10.1088/1751-8121/aa714b}{{\em J. Phys.} {\bfseries
  A50} no.~25, (2017) 254001},
\href{http://arxiv.org/abs/1612.06236}{{\ttfamily arXiv:1612.06236 [hep-th]}}.

\bibitem{Feigin:2017}
B.~Feigin, M.~Jimbo, and E.~Mukhin, ``{Integrals of motion from quantum
  toroidal algebras},'' {\em J. Phys.} {\bfseries A50} no.~46, (2017) 464001,
  \href{http://arxiv.org/abs/1705.07984}{{\ttfamily arXiv:1705.07984
  [math.QA]}}.

\bibitem{Hutsalyuk:2016ndz}
A.~Hutsalyuk, A.~Liashyk, S.~Z. Pakuliak, E.~Ragoucy, and N.~A. Slavnov,
  ``{Scalar products of Bethe vectors in models with $\mathfrak{gl}(2|1)$
  symmetry 1. Super-analog of Reshetikhin formula},''
  \href{http://dx.doi.org/10.1088/1751-8113/49/45/454005}{{\em J. Phys.}
  {\bfseries A49} no.~45, (2016) 454005},
\href{http://arxiv.org/abs/1605.09189}{{\ttfamily arXiv:1605.09189 [math-ph]}}.

\bibitem{Hutsalyuk:2016yii}
A.~Hutsalyuk, A.~Liashyk, S.~Z. Pakuliak, E.~Ragoucy, and N.~A. Slavnov,
  ``{Scalar products of Bethe vectors in models with
  $\mathfrak{g}\mathfrak{l}(2|1)$ symmetry 2. Determinant representation},''
  \href{http://dx.doi.org/10.1088/1751-8121/50/3/034004}{{\em J. Phys.}
  {\bfseries A50} no.~3, (2017) 034004},
\href{http://arxiv.org/abs/1606.03573}{{\ttfamily arXiv:1606.03573 [math-ph]}}.

\bibitem{Kuniba:2016iam}
A.~Kuniba and M.~Okado, ``{Matrix product formula for ${{U}_{q}}(A_{n}^{(1)})$
  -zero range process},''
  \href{http://dx.doi.org/10.1088/1751-8121/50/4/044001}{{\em J. Phys.}
  {\bfseries A50} no.~4, (2017) 044001},
\href{http://arxiv.org/abs/1608.02779}{{\ttfamily arXiv:1608.02779 [math.QA]}}.

\bibitem{Marboe:2016yyn}
C.~Marboe and D.~Volin, ``{Fast analytic solver of rational Bethe equations},''
  \href{http://dx.doi.org/10.1088/1751-8121/aa6b88}{{\em J. Phys.} {\bfseries
  A50} no.~20, (2017) 204002},
\href{http://arxiv.org/abs/1608.06504}{{\ttfamily arXiv:1608.06504 [math-ph]}}.

\bibitem{Fioravanti:2016bmi}
D.~Fioravanti and R.~I. Nepomechie, ``{An inhomogeneous Lax representation for
  the Hirota equation},''
  \href{http://dx.doi.org/10.1088/1751-8121/aa5303}{{\em J. Phys.} {\bfseries
  A50} no.~5, (2017) 054001},
\href{http://arxiv.org/abs/1609.06761}{{\ttfamily arXiv:1609.06761 [math-ph]}}.

\bibitem{Ahmed:2017mqq}
I.~Ahmed, R.~I. Nepomechie, and C.~Wang, ``{Quantum group symmetries and
  completeness for $A_{2n}^{(2)}$ open spin chains},''
  \href{http://dx.doi.org/10.1088/1751-8121/aa7606}{{\em J. Phys.} {\bfseries
  A50} no.~28, (2017) 284002},
\href{http://arxiv.org/abs/1702.01482}{{\ttfamily arXiv:1702.01482 [math-ph]}}.

\bibitem{Stronks:2016nml}
L.~Stronks, J.~van~de Leur, and D.~Schuricht, ``{On rational R-matrices with
  adjoint $SU(n)$ symmetry},''
  \href{http://dx.doi.org/10.1088/1751-8113/49/44/444001}{{\em J. Phys.}
  {\bfseries A49} no.~44, (2016) 444001},
\href{http://arxiv.org/abs/1606.02516}{{\ttfamily arXiv:1606.02516 [math-ph]}}.

\bibitem{Belletete:2017gwt}
J.~Bellette, A.~M. Gainutdinov, J.~L. Jacobsen, H.~Saleur, and R.~Vasseur,
  ``{On the correspondence between boundary and bulk lattice models and
  (logarithmic) conformal field theories},'' {\em J. Phys.} {\bfseries A50}
  no.~48, (2017) 484002,
\href{http://arxiv.org/abs/1705.07769}{{\ttfamily arXiv:1705.07769 [hep-th]}}.

\bibitem{Ymai:2017yau}
L.~H. Ymai, A.~P. Tonel, A.~Foerster, and J.~Links, ``{Quantum integrable
  multi-well tunneling models},''
  \href{http://dx.doi.org/10.1088/1751-8121/aa7227}{{\em J. Phys.} {\bfseries
  A50} no.~26, (2017) 264001},
\href{http://arxiv.org/abs/1606.00816}{{\ttfamily arXiv:1606.00816 [math-ph]}}.

\bibitem{Bogoliubov:2017}
N.~Bogoliubov, I.~Ermakov, and A.~Rybin, ``{Time evolution of the atomic
  inversion for the generalized Tavis-Cummings model - QIM approach},'' {\em J.
  Phys.} {\bfseries A50} no.~46, (2017) 464003,
  \href{http://arxiv.org/abs/1702.03740}{{\ttfamily arXiv:1702.03740
  [quant-ph]}}.

\bibitem{Zhang:2016kas}
X.~Zhang, J.~Cao, W.-L. Yang, K.~Shi, and Y.~Wang, ``{Exact solution of the
  relativistic quantum Toda chain},''
  \href{http://dx.doi.org/10.1088/1751-8121/aa5c0f}{{\em J. Phys.} {\bfseries
  A50} no.~12, (2017) 124003},
\href{http://arxiv.org/abs/1609.07385}{{\ttfamily arXiv:1609.07385 [math-ph]}}.

\bibitem{Jones:2017}
V.~F. Jones, ``{Scale invariant transfer matrices and Hamiltionians},''
  \href{http://arxiv.org/abs/1706.00515}{{\ttfamily arXiv:1706.00515
  [math.OA]}}.

\bibitem{Derkachov:2016dhc}
S.~E. Derkachov and A.~N. Manashov, ``{Spin chains and GustafsonÕs
  integrals},'' \href{http://dx.doi.org/10.1088/1751-8121/aa749a}{{\em J.
  Phys.} {\bfseries A50} no.~29, (2017) 294006},
\href{http://arxiv.org/abs/1611.09593}{{\ttfamily arXiv:1611.09593 [math-ph]}}.

\bibitem{Derkachov:2016ucn}
S.~E. Derkachov, A.~N. Manashov, and P.~A. Valinevich, ``{Gustafson integrals
  for $SL(2, \mathbb{C})$ spin magnet},''
  \href{http://dx.doi.org/10.1088/1751-8121/aa7480}{{\em J. Phys.} {\bfseries
  A50} no.~29, (2017) 294007},
\href{http://arxiv.org/abs/1612.00727}{{\ttfamily arXiv:1612.00727 [math-ph]}}.

\bibitem{Kitanine:2016pvg}
N.~Kitanine, J.~M. Maillet, G.~Niccoli, and V.~Terras, ``{The open XXX spin
  chain in the SoV framework: scalar product of separate states},''
  \href{http://dx.doi.org/10.1088/1751-8121/aa6cc9}{{\em J. Phys.} {\bfseries
  A50} no.~22, (2017) 224001},
\href{http://arxiv.org/abs/1606.06917}{{\ttfamily arXiv:1606.06917 [math-ph]}}.

\bibitem{Dugave:2016hep}
M.~Dugave, F.~Gohmann, K.~K. Kozlowski, and J.~Suzuki, ``{Thermal form factor
  approach to the ground-state correlation functions of the XXZ chain in the
  antiferromagnetic massive regime},''
  \href{http://dx.doi.org/10.1088/1751-8113/49/39/394001}{{\em J. Phys.}
  {\bfseries A49} no.~39, (2016) 394001},
\href{http://arxiv.org/abs/1605.07968}{{\ttfamily arXiv:1605.07968
  [cond-mat.stat-mech]}}.

\bibitem{Groha:2017wpn}
S.~Groha and F.~H.~L. Essler, ``{Spinon decay in the spin-1/2 Heisenberg chain
  with weak next nearest neighbour exchange},''
  \href{http://dx.doi.org/10.1088/1751-8121/aa7d41}{{\em J. Phys.} {\bfseries
  A50} no.~33, (2017) 334002},
\href{http://arxiv.org/abs/1702.06550}{{\ttfamily arXiv:1702.06550
  [cond-mat]}}.

\bibitem{Frahm:2017qfh}
H.~Frahm and K.~Hobu$\ss$, ``{Spectral flow for an integrable staggered
  superspin chain},'' \href{http://dx.doi.org/10.1088/1751-8121/aa77e7}{{\em J.
  Phys.} {\bfseries A50} no.~29, (2017) 294002},
\href{http://arxiv.org/abs/1703.08054}{{\ttfamily arXiv:1703.08054
  [cond-mat.stat-mech]}}.

\bibitem{Pearce:2016vgu}
P.~A. Pearce and A.~Vittorini-Orgeas, ``{Yang-Baxter Solution of Dimers as a
  Free-Fermion Six-Vertex Model},''
  \href{http://dx.doi.org/10.1088/1751-8121/aa86bc}{{\em J. Phys.} {\bfseries
  A50} no.~43, (2017) 434001},
\href{http://arxiv.org/abs/1612.09477}{{\ttfamily arXiv:1612.09477 [math-ph]}}.

\bibitem{Burban:2017}
I.~Burban, L.~Galinat, and A.~Stolin, ``{Simple vector bundles on a nodal
  Weierstrass cubic and quasi-trigonometric solutions of the classical
  Yang-Baxter equation},'' {\em J. Phys.} {\bfseries A50} (2017) 454002,
  \href{http://arxiv.org/abs/1704.07202}{{\ttfamily arXiv:1704.07202
  [math.AG]}}.

\bibitem{Beisert:2017xqx}
N.~Beisert, R.~Hecht, and B.~Hoare, ``{Maximally extended $\boldsymbol
  {\mathfrak{sl}(2\vert 2)}$ , q-deformed $\boldsymbol{\mathfrak{d}{(2,
  1;\epsilon)}}$ and 3D kappa-Poincar\'e},''
  \href{http://dx.doi.org/10.1088/1751-8121/aa7a2f}{{\em J. Phys.} {\bfseries
  A50} no.~31, (2017) 314003},
\href{http://arxiv.org/abs/1704.05093}{{\ttfamily arXiv:1704.05093 [math-ph]}}.

\bibitem{Avan:2017xzz}
J.~Avan, L.~Frappat, and E.~Ragoucy, ``{Dynamical centers for the elliptic
  quantum algebra ${\mathcal{B}}_{q,\lambda}(\hat{gl}_2)_c$},''
  \href{http://dx.doi.org/10.1088/1751-8121/aa85b2}{{\em J. Phys.} {\bfseries
  A50} no.~39, (2017) 394002},
\href{http://arxiv.org/abs/1703.05223}{{\ttfamily arXiv:1703.05223 [math-ph]}}.

\bibitem{Reshetikhin:2017}
N.~Reshetikhin, ``{Semiclassical geometry of integrable systems},''.

\bibitem{MR0244547}
P.~P. Kulish, ``The inverse scattering problem for {S}chr\"odinger's equation
  on the axis,'' {\em Mat. Zametki} {\bfseries 4} (1968) 677--684. MR0244547

\bibitem{more1}
P.~P. Kulish and L.~D. Faddeev, ``Asymptotic conditions and infrared
  divergences in {QED},'' {\em Teoret. Mat. Fiz.} {\bfseries 4} (1970)
  153--170. Theor. Math. Phys. 4 (1970) 745-757.

\bibitem{ISI:A1970G610100018}
M.~Y. Amusia, M.~P. Kazachkov, and P.~P. Kulish, ``On collective spectrum of a
  spin density wave electron gas,''
  \href{http://dx.doi.org/{10.1016/0375-9601(70)90065-4}}{{\em {Physics
  Letters}} {\bfseries {A 32}} no.~{1}, ({1970}) {27+}}. ISI:A1970G610100018

\bibitem{more2}
P.~P. Kulish, ``Asymptotic conditions and infrared divergences in quantum field
  theory,'' 1971.
\newblock Autoreferat of Ph.D. Thesis (in Russian), 12pp, Leningrad State
  University.

\bibitem{Kulish1971}
P.~P. Kulish, ``Asymptotic states of massive particles interacting with the
  gravitational field,'' \href{http://dx.doi.org/10.1007/BF01037574}{{\em
  Theoretical and Mathematical Physics} {\bfseries 6} no.~1, (Jan, 1971)
  18--23}. \url{https://doi.org/10.1007/BF01037574}.

\bibitem{more3}
P.~P. Kulish, ``Conservation laws for the sine-{G}ordon equation.''.
  (Serpukhov, IFVE), IFVE-74-155, Dec. 1974, 7pp.

\bibitem{MR0468719}
I.~Y. Aref'eva and P.~P. Kulish, ``Representations of the canonical commutation
  relations in the limit of infinite volume,'' {\em Teoret. Mat. Fiz.}
  {\bfseries 17} (1973) 3--18. MR0468719

\bibitem{Gerdzhikov1974a}
V.~S. Gerdzhikov and P.~P. Kulish, ``Low-energy for photons and infrared
  divergences,'' \href{http://dx.doi.org/10.1007/BF01036923}{{\em Theoretical
  and Mathematical Physics} {\bfseries 18} no.~1, (Jan, 1974) 36--38}.
  \url{https://doi.org/10.1007/BF01036923}.

\bibitem{Gerdzhikov1974b}
V.~S. Gerdzhikov and P.~P. Kulish, ``Low-energy structure of the feynman s
  matrix,'' \href{http://dx.doi.org/10.1007/BF01035553}{{\em Theoretical and
  Mathematical Physics} {\bfseries 21} no.~2, (Nov, 1974) 1065--1073}.
  \url{https://doi.org/10.1007/BF01035553}.

\bibitem{MR0522299}
E.~V. Damaskinsky and P.~P. Kulish, ``The current group of the {T}hirring model
  and its representations,'' {\em Izv. Vys\v s. U\v cebn. Zaved. Fizika}
  no.~11, (1975) 57--62. MR0522299

\bibitem{ISI:A1975AL15700014}
V.~E. Korepin, P.~P. Kulish, and L.~D. Faddeev, ``Soliton quantization,'' {\em
  JETP Letters} {\bfseries {21}} no.~{5}, ({1975}) {138--139}. ISI:A1975AL15700014

\bibitem{more5}
V.~E. Korepin, P.~P. Kulish, and L.~D. Faddeev, ``Quantization of solitons (in
  {R}ussian).''. Erevan 1975, Proceedings, Problems of Elementary Particle
  Physics Vol. 2, Erevan 1976, 458-464.

\bibitem{MR0449295}
P.~P. Kulish and E.~R. Nisimov, ``Anomalies of quantum currents in exactly
  solvable models,'' {\em Teoret. Mat. Fiz.} {\bfseries 29} no.~2, (1976)
  161--170. MR0449295

\bibitem{more6}
P.~P. Kulish and E.~R. Nisimov, ``Conservation laws in the quantum theory
  $\cos\phi$ in two-dimensions and in the massive {T}hirring model,'' {\em
  Pisma Zh. Eksperim. I Teor. Fiz.} {\bfseries 24} (1976) 247--250.

\bibitem{more7}
I.~Y. Aref'eva, P.~P. Kulish, E.~R. Nisimov, and S.~J. Pacheva, ``Infinite set
  of conservation laws of the quantum chiral field in two-dimensional
  space-time.''. (Steklov Math. Inst., Leningrad), LOMI E-1-1978, Oct 1977,
  29pp.

\bibitem{MR0462300}
P.~P. Kulish, S.~V. Manakov, and L.~D. Faddeev, ``Comparison of the exact
  quantum and quasiclassical results for the nonlinear {S}chr\"odinger
  equation,'' {\em Teoret. Mat. Fiz.} {\bfseries 28} no.~1, (1976) 
  38--45. MR0462300

\bibitem{MR0468929}
P.~P. Kulish, ``Factorization of the classical and the quantum {$S$}-matrix,
  and conservation laws,'' {\em Teoret. Mat. Fiz.} {\bfseries 26} no.~2, (1976)
  198--205. MR0468929

\bibitem{MR0456130}
P.~P. Kulish, ``Conservation laws for a string in a static field,'' {\em
  Teoret. Mat. Fiz.} {\bfseries 33} no.~2, (1977) 272--275. MR0456130

\bibitem{MR518440}
L.~D. Faddeev and P.~P. Kulish, ``Quantization of particle-like solutions in
  field theory,'' in {\em Mathematical problems in theoretical physics ({P}roc.
  {I}nternat. {C}onf., {U}niv. {R}ome, {R}ome, 1977)}, vol.~80 of {\em Lecture
  Notes in Phys.}, pp.~270--278.
\newblock Springer, Berlin-New York, 1978. MR518440

\bibitem{MR520578}
V.~S. Gerdzhikov and P.~P. Kulish, ``Completely integrable {H}amiltonian
  systems connected with a nonselfadjoint {D}irac operator,'' {\em Bulgar. J.
  Phys.} {\bfseries 5} no.~4, (1978) 337--348. MR520578

\bibitem{MR521014}
P.~P. Kulish, ``Factorization of scattering characteristics and integrals of
  motion,'' in {\em Nonlinear evolution equations solvable by the spectral
  transform ({I}nternat. {S}ympos., {A}ccad. {L}incei, {R}ome, 1977)}, vol.~26
  of {\em Res. Notes in Math.}, pp.~252--257.
\newblock Pitman, Boston, Mass.-London, 1978. MR521014

\bibitem{MR541695}
A.~G. Izergin and P.~P. Kulish, ``A massive {T}hirring model with field values
  in the {G}rassmann algebra,'' {\em Zap. Nauchn. Sem. Leningrad. Otdel. Mat.
  Inst. Steklov. (LOMI)} {\bfseries 77} (1978) 76--83, 230. Questions in
  quantum field theory and statistical physics. MR541695

\bibitem{more8}
A.~G. Izergin and P.~P. Kulish, ``On the inverse scattering method for the
  classical massive {T}hirring model with anticommuting variables,'' {\em Lett.
  Math. Phys.} {\bfseries 2} (1978) 297--302.

\bibitem{more9}
M.~Chaichian and P.~P. Kulish, ``On the method of inverse scattering problem
  and {B}\"acklund transformations for supersymmetric equations,'' {\em Phys.
  Lett.} {\bfseries B 78} (1978) 413.
  
\bibitem{MR541697}
P.~P. Kulish, ``Infrared divergences of a quantized gravitational field,'' {\em
  Zap. Nauchn. Sem. Leningrad. Otdel. Mat. Inst. Steklov. (LOMI)} {\bfseries
  77} (1978) 106--123, 230--231. Questions in quantum field theory and
  statistical physics. MR541697

\bibitem{MR541698}
P.~P. Kulish and V.~N. Popov, ``Infrared asymptotic behavior of the {G}reen
  function of massive particles in a charge-symmetric model,'' {\em Zap.
  Nauchn. Sem. Leningrad. Otdel. Mat. Inst. Steklov. (LOMI)} {\bfseries 77}
  (1978) 124--133, 231. Questions in quantum field theory and statistical
  physics. MR541698

\bibitem{MR541699}
P.~P. Kulish and A.~G. Reyman, ``The hierarchy of symplectic forms for the
  {S}chr\"odinger equation and for the {D}irac equation,'' {\em Zap. Nauchn.
  Sem. Leningrad. Otdel. Mat. Inst. Steklov. (LOMI)} {\bfseries 77} (1978)
  134--147, 231. Questions in quantum field theory and statistical 
  physics. MR541699

\bibitem{MR536467}
V.~S. Gerdzhikov and P.~P. Kulish, ``Derivation of the {B}\"acklund
  transformation in the formalism of the inverse scattering problem,'' {\em
  Teoret. Mat. Fiz.} {\bfseries 39} no.~1, (1979) 69--74. MR536467

\bibitem{more10}
P.~P. Kulish and E.~K. Sklyanin, ``Solutions of the {Y}ang-{B}axter equation,''
  {\em Zap. Nauch. Sem. LOMI} {\bfseries 95} (1980) 129--160. Jour. Sov. Math.
  19 No 5 (1982) 1596.

\bibitem{MR588129}
P.~P. Kulish and E.~K. Sklyanin, ``Quantum inverse scattering method and the
  {H}eisenberg ferromagnet,''
  \href{http://dx.doi.org/10.1016/0375-9601(79)90365-7}{{\em Phys. Lett. A}
  {\bfseries 70} no.~5-6, (1979) 461--463}.
  \url{http://dx.doi.org/10.1016/0375-9601(79)90365-7}. MR588129

\bibitem{more11}
P.~P. Kulish, ``Generalized {B}ethe ansatz and quantum inverse problem method
  (in {R}ussian).''. (Steklov Math. Inst., Leningrad), LOMI-P-3-79, April 1979,
  16pp.

\bibitem{MR579477}
P.~P. Kulish, ``Generating operators for integrable nonlinear evolution
  equations,'' {\em Zap. Nauchn. Sem. Leningrad. Otdel. Mat. Inst. Steklov.
  (LOMI)} {\bfseries 96} (1980) 105--112, 307--308. Boundary value problems of
  mathematical physics and related questions in the theory of 
  functions, 12. MR579477

\bibitem{MR582629}
V.~S. Gerdjikov, P.~P. Kulish, and M.~I. Ivanov, ``Classical and quantum
  aspects of the inverse scattering method,'' in {\em Mathematical problems in
  theoretical physics ({P}roc. {I}nternat. {C}onf. {M}ath. {P}hys., {L}ausanne,
  1979)}, vol.~116 of {\em Lecture Notes in Phys.}, pp.~244--248.
\newblock Springer, Berlin-New York, 1980. MR582629

\bibitem{MR587299}
A.~G. Izergin and P.~P. Kulish, ``Inverse problem for systems with
  anticommuting variables and the massive {T}hirring model,'' {\em Teoret. Mat.
  Fiz.} {\bfseries 44} no.~2, (1980) 189--193. MR587299

\bibitem{MR596207}
V.~S. Gerdzhikov, M.~I. Ivanov, and P.~P. Kulish, ``Quadratic pencils and
  nonlinear equations,'' {\em Teoret. Mat. Fiz.} {\bfseries 44} no.~3, (1980)
  342--357. MR596207

\bibitem{MR612869}
P.~P. Kulish, ``Multicomponent nonlinear {S}chr\"odinger equation with
  grading,'' {\em Dokl. Akad. Nauk SSSR} {\bfseries 255} no.~2, (1980)
  323--326. MR612869

\bibitem{more12}
V.~S. Gerdzhikov, M.~I. Ivanov, and P.~P. Kulish, ``Complete integrability of
  the difference evolution equations.''. JINR-E2-80-882, Dec 1980, 20pp,
  submitted to Rept. Math. Phys.

\bibitem{more13}
L.~D. Faddeev and P.~P. Kulish, ``Development of the quantum inverse problem
  method (abstract only).''. Chania 1980, Proceedings, Nonlinear Evolution
  Equations and Dynamical Systems, 21.

\bibitem{MR612953}
P.~P. Kulish and S.~A. Cypljaev, ``Supersymmetric {${\rm cos}\,\Phi _{2}$}\
  model and the inverse problem method,'' {\em Teoret. Mat. Fiz.} {\bfseries
  46} no.~2, (1981) 172--186. MR612953

\bibitem{MR623924}
V.~S. Gerdzhikov and P.~P. Kulish, ``Expansion in ``squares''\ of
  eigenfunctions of a matrix linear system,'' {\em Zap. Nauchn. Sem. Leningrad.
  Otdel. Mat. Inst. Steklov. (LOMI)} {\bfseries 101} (1981) 46--63, 206.
  Questions in quantum field theory and statistical physics, 2. MR623924

\bibitem{MR623928}
P.~P. Kulish and N.~Y. Reshetihin, ``Quantum linear problem for the
  sine-{G}ordon equation and higher representations,'' {\em Zap. Nauchn. Sem.
  Leningrad. Otdel. Mat. Inst. Steklov. (LOMI)} {\bfseries 101} (1981)
  101--110, 207. Questions in quantum field theory and statistical 
  physics, 2. MR623928

\bibitem{MR624761}
P.~P. Kulish, ``Quantum difference nonlinear {S}chr\"odinger equation,''
  \href{http://dx.doi.org/10.1007/BF00420698}{{\em Lett. Math. Phys.}
  {\bfseries 5} no.~3, (1981) 191--197}.
  \url{http://dx.doi.org/10.1007/BF00420698}. MR624761

\bibitem{MR628604}
P.~P. Kulish and E.~K. Sklyanin, ``{${\rm O}(N)$}-invariant nonlinear
  {S}chr\"odinger equation---a new completely integrable system,''
  \href{http://dx.doi.org/10.1016/0375-9601(81)90205-X}{{\em Phys. Lett. A}
  {\bfseries 84} no.~7, (1981) 349--352}.
  \url{http://dx.doi.org/10.1016/0375-9601(81)90205-X}. MR628604

\bibitem{MR629116}
P.~P. Kulish, ``Realization of the {Z}amolodchikov-{F}addeev algebra,'' {\em
  Zap. Nauchn. Sem. Leningrad. Otdel. Mat. Inst. Steklov. (LOMI)} {\bfseries
  109} (1981) 83--92, 181, 183. Differential geometry, Lie groups and
  mechanics, IV. MR629116

\bibitem{MR631186}
V.~S. Gerdjikov and P.~P. Kulish, ``The generating operator for the {$n\times
  n$}\ linear system,''
  \href{http://dx.doi.org/10.1016/0167-2789(81)90039-7}{{\em Phys. D}
  {\bfseries 3} no.~3, (1981) 549--564}.
  \url{http://dx.doi.org/10.1016/0167-2789(81)90039-7}. MR631186

\bibitem{more14}
P.~P. Kulish and N.~Y. Reshetikhin, ``Generalized {H}eisenberg ferromagnet and
  the {G}ross-{N}eveu model,'' {\em Sov. Phys. JETP} no.~53 (1), (1981)
  108--114. Zh. Eksp. Fiz. 80 (1981), 214-228.

\bibitem{MR649704}
P.~P. Kulish, N.~Y. Reshetikhin, and E.~K. Sklyanin, ``Yang-{B}axter equations
  and representation theory. {I},''
  \href{http://dx.doi.org/10.1007/BF02285311}{{\em Lett. Math. Phys.}
  {\bfseries 5} no.~5, (1981) 393--403}.
  \url{http://dx.doi.org/10.1007/BF02285311}. MR649704

\bibitem{MR681524}
P.~P. Kulish and S.~A. Tsyplyaev, ``Calculation of the {P}oisson brackets of
  scattering data in the inverse problem method,'' in {\em Generalized
  functions and their applications in mathematical physics ({M}oscow, 1980)},
  pp.~309--315.
\newblock Akad. Nauk SSSR, Vychisl. Tsentr, Moscow, 1981. MR681524

\bibitem{more15}
P.~P. Kulish, ``Quantum inverse problem method and exactly solvable models of
  statistical physics.''. Proceedings, II International Symposium on Selected
  Topics in Statistical Mechanics, Dubna (1981) 147-157.

\bibitem{ISI:A1981MH77800015}
P.~P. Kulish, ``Classical and quantum inverse problem method and generalized
  {B}ethe ansatz,''
  \href{http://dx.doi.org/{10.1016/0167-2789(81)90130-5}}{{\em Physica D}
  {\bfseries {3}} no.~{1-2}, ({1981}) {246--257}}. ISI:A1981MH77800015

\bibitem{MR660077}
P.~P. Kulish, ``Action-angle variables for a multicomponent nonlinear
  {S}chr\"odinger equation,'' {\em Zap. Nauchn. Sem. Leningrad. Otdel. Mat.
  Inst. Steklov. (LOMI)} {\bfseries 115} (1982) 126--136, 308. Boundary value
  problems of mathematical physics and related questions in the theory of
  functions, 14. MR660077

\bibitem{ISI:A1982NN06800015}
P.~P. Kulish, ``Correction,'' {\em {Physics Letters A}} {\bfseries {88}}
  no.~{9}, ({1982}) {491}. ISI:A1982NN06800015

\bibitem{MR664161}
P.~P. Kulish and F.~A. Smirnov, ``Quantum inverse problem and {G}reen functions
  for the {H}eisenberg ferromagnet,''
  \href{http://dx.doi.org/10.1016/0375-9601(82)90055-X}{{\em Phys. Lett. A}
  {\bfseries 90} no.~1-2, (1982) 74--78}.
  \url{http://dx.doi.org/10.1016/0375-9601(82)90055-X}. MR664161

\bibitem{MR671263}
P.~P. Kulish and E.~K. Sklyanin, ``Quantum spectral transform method. {R}ecent
  developments,'' vol.~151 of {\em Lecture Notes in Phys.}, pp.~61--119.
\newblock Springer, Berlin-New York, 1982. MR671263

\bibitem{MR674783}
N.~V. Borisov and P.~P. Kulish, ``Path integral in superspace for a
  relativistic spinor particle in an external gauge field,'' {\em Teoret. Mat.
  Fiz.} {\bfseries 51} no.~3, (1982) 335--343. MR674783

\bibitem{more16}
P.~P. Kulish, ``Path integral in superspace (in {R}ussian).''. Sukhumi 1982,
  Proceedings, Quarks-82, 166-170.

\bibitem{MR701555}
P.~P. Kulish and N.~Y. Reshetikhin, ``On {${\rm GL}_{3}$}-invariant solutions
  of the {Y}ang-{B}axter equation and associated quantum systems,'' {\em Zap.
  Nauchn. Sem. Leningrad. Otdel. Mat. Inst. Steklov. (LOMI)} {\bfseries 120}
  (1982) 92--121. Questions in quantum field theory and statistical 
  physics, 3. MR701555

\bibitem{MR701556}
P.~P. Kulish and S.~A. Tsyplyaev, ``Complete integrability of the
  supersymmetric model {$({\rm cos}\,\varphi )_{2}$},'' {\em Zap. Nauchn. Sem.
  Leningrad. Otdel. Mat. Inst. Steklov. (LOMI)} {\bfseries 120} (1982)
  122--135. Questions in quantum field theory and statistical 
  physics, 3. MR701556

\bibitem{MR697242}
P.~P. Kulish and A.~G. Reyman, ``Hamiltonian structure of polynomial bundles,''
  {\em Zap. Nauchn. Sem. Leningrad. Otdel. Mat. Inst. Steklov. (LOMI)}
  {\bfseries 123} (1983) 67--76. Differential geometry, Lie groups and
  mechanics, V. MR697242

\bibitem{MR709476}
A.~P. Fordy and P.~P. Kulish, ``Nonlinear {S}chr\"odinger equations and simple
  {L}ie algebras,'' {\em Comm. Math. Phys.} {\bfseries 89} no.~3, (1983)
  427--443. \url{http://projecteuclid.org/euclid.cmp/1103922818}. MR709476

\bibitem{MR718678}
V.~S. Gerdzhikov and P.~P. Kulish, ``The multicomponent nonlinear
  {S}chr\"odinger equation in the case of nonzero boundary conditions,'' {\em
  Zap. Nauchn. Sem. Leningrad. Otdel. Mat. Inst. Steklov. (LOMI)} {\bfseries
  131} (1983) 34--46. Questions in quantum field theory and statistical
  physics, 4. MR718678

\bibitem{more17}
P.~P. Kulish, ``Quantum inverse scattering method for multicomponent
  systems.''. Autoreferat of Doctor of Science Thesis, Steklov Mathematical
  Institute, Moscow 1983, 21pp.

\bibitem{MR727044}
P.~P. Kulish and N.~Y. Reshetikhin, ``Diagonalisation of {${\rm GL}(N)$}\
  invariant transfer matrices and quantum {$N$}-wave system ({L}ee model),''
  {\em J. Phys. A} {\bfseries 16} no.~16, (1983) L591--L596.
  \url{http://stacks.iop.org/0305-4470/16/L591}. MR727044

\bibitem{MR728883}
V.~S. Gerdzhikov, M.~I. Ivanov, and P.~P. Kulish, ``Expansions over the
  ``squared''\ solutions and difference evolution equations,''
  \href{http://dx.doi.org/10.1063/1.525994}{{\em J. Math. Phys.} {\bfseries 25}
  no.~1, (1984) 25--34}. \url{http://dx.doi.org/10.1063/1.525994}. MR728883

\bibitem{MR742155}
P.~P. Kulish and N.~Y. Reshetikhin, ``Integrable fermion chiral models
  connected with classical {L}ie algebras,'' {\em Zap. Nauchn. Sem. Leningrad.
  Otdel. Mat. Inst. Steklov. (LOMI)} {\bfseries 133} (1984) 146--159.
  Differential geometry, Lie groups and mechanics, VI.

\bibitem{MR797004}
P.~P. Kulish, ``Quantum {${\rm osp}$}-invariant nonlinear {S}chr\"odinger
  equation,'' \href{http://dx.doi.org/10.1007/BF00704591}{{\em Lett. Math.
  Phys.} {\bfseries 10} no.~1, (1985) 87--93}.
  \url{http://dx.doi.org/10.1007/BF00704591}. MR797004

\bibitem{MR825250}
P.~P. Kulish, ``Quantum spectral transform,'' in {\em Critical phenomena
  ({B}ra\c sov, 1983)}, vol.~11 of {\em Progr. Phys.}, pp.~351--374.
\newblock Birkh\"auser Boston, Boston, MA, 1985. MR825250

\bibitem{MR857958}
S.~I. Alishauskas and P.~P. Kulish, ``Spectral expansion of {${\rm
  SU}(3)$}-invariant solutions of the {Y}ang-{B}axter equation,'' {\em Zap.
  Nauchn. Sem. Leningrad. Otdel. Mat. Inst. Steklov. (LOMI)} {\bfseries 145}
  no.~Voprosy Kvant. Teor. Polya i Statist. Fiz. 5, (1985) 3--21, 
  189, 193. MR857958

\bibitem{MR857967}
P.~P. Kulish, ``Integrable graded magnets,'' {\em Zap. Nauchn. Sem. Leningrad.
  Otdel. Mat. Inst. Steklov. (LOMI)} {\bfseries 145} no.~Voprosy Kvant. Teor.
  Polya i Statist. Fiz. 5, (1985) 140--163, 191, 195. MR857967

\bibitem{MR890151}
P.~P. Kulish, ``Solitons and representation theory,'' in {\em Problems of
  nonlinear and turbulent processes in physics, {P}art 1 ({R}ussian) ({K}iev,
  1983)}, pp.~53--59.
\newblock ``Naukova Dumka'', Kiev, 1985. MR890151

\bibitem{ISI:A1985ACK0600011}
P.~P. Kulish and F.~A. Smirnov, ``Anisotropic {H}eisenberg-ferromagnet with a
  ground-state of the domain-wall type,'' {\em Journal of Physics C-Solid State
  Physics}. ISI:A1985ACK0600011

\bibitem{MR838342}
P.~P. Kulish, ``Quantum nonlinear wave interaction system,''
  \href{http://dx.doi.org/10.1016/0167-2789(86)90197-1}{{\em Phys. D}
  {\bfseries 18} no.~1-3, (1986) 360--364}.
  \url{http://dx.doi.org/10.1016/0167-2789(86)90197-1}. Solitons and coherent
  structures (Santa Barbara, Calif., 1985). MR838342

\bibitem{MR861261}
P.~P. Kulish and F.~A. Smirnov, ``Equations of the inverse problem of the
  quantum three-wave system,'' \href{http://dx.doi.org/10.1007/BF01099191}{{\em
  Zap. Nauchn. Sem. Leningrad. Otdel. Mat. Inst. Steklov. (LOMI)} {\bfseries
  150} no.~Voprosy Kvant. Teor. Polya i Statist. Fiz. 6, (1986) 53--69, 220}.
  \url{http://dx.doi.org/10.1007/BF01099191}. MR861261
  
\bibitem{MR869580}
P.~P. Kulish, ``An analogue of the {K}orteweg-de {V}ries equation for the
  superconformal algebra,'' \href{http://dx.doi.org/10.1007/BF01247091}{{\em
  Zap. Nauchn. Sem. Leningrad. Otdel. Mat. Inst. Steklov. (LOMI)} {\bfseries
  155} no.~Differentsial'naya Geometriya, Gruppy Li i Mekh. VIII, (1986)
  142--149, 194}. \url{http://dx.doi.org/10.1007/BF01247091}. MR869580

\bibitem{MR919751}
P.~P. Kulish, ``Integrable models of field theory and simple {L}ie algebras and
  superalgebras,'' in {\em Group theoretical methods in physics, {V}ol.\ {I}
  ({Y}urmala, 1985)}, pp.~299--305.
\newblock VNU Sci. Press, Utrecht, 1986. MR919751

\bibitem{MR946845}
P.~P. Kulish, ``Integrable models in field theory, and simple {L}ie algebras
  and {L}ie superalgebras,'' in {\em Group-theoretic methods in physics,
  {V}ol.\ 1 ({R}ussian) ({J}\=urmala, 1985)}, pp.~412--417.
\newblock ``Nauka'', Moscow, 1986. MR946845

\bibitem{more18}
P.~P. Kulish, ``Algebraic and {H}amiltonian methods in the theory of nonabelian
  anomalies (abstract only).''. Espoo 1986, Proceedings, Topological and
  geometrical methods in field theory, 341.

\bibitem{MR874098}
M.~Chaichian and P.~P. Kulish, ``Superconformal algebras and their relation to
  integrable nonlinear systems,''
  \href{http://dx.doi.org/10.1016/0370-2693(87)90432-1}{{\em Phys. Lett. B}
  {\bfseries 183} no.~2, (1987) 169--174}.
  \url{http://dx.doi.org/10.1016/0370-2693(87)90432-1}. MR874098

\bibitem{MR918753}
P.~P. Kulish and V.~D. Lipovski\u\i, ``Hamiltonian interpretation of the
  inverse problem method for the {D}avey-{S}tewartson equation,''
  \href{http://dx.doi.org/10.1007/BF01096091}{{\em Zap. Nauchn. Sem. Leningrad.
  Otdel. Mat. Inst. Steklov. (LOMI)} {\bfseries 161} no.~Vopr. Kvant. Teor.
  Polya i Statist. Fiz. 7, (1987) 54--71, 176, 179}.
  \url{http://dx.doi.org/10.1007/BF01096091}. MR918753

\bibitem{MR939802}
P.~P. Kulish and M.~A. Sokolov, ``On scattering theory in the massless {L}ee
  model,'' {\em Teoret. Mat. Fiz.} {\bfseries 73} no.~1, (1987) 
  149--153. MR939802

\bibitem{MR947336}
P.~P. Kulish, V.~D. Lipovski\u\i, and A.~V. Shirokov, ``Scattering data for the
  nonstationary {D}irac equation,''
  \href{http://dx.doi.org/10.1007/BF01840430}{{\em Zap. Nauchn. Sem. Leningrad.
  Otdel. Mat. Inst. Steklov. (LOMI)} {\bfseries 164} no.~Differentsial'naya
  Geom. Gruppy Li i Mekh. IX, (1987) 169--175, 199}.
  \url{http://dx.doi.org/10.1007/BF01840430}. MR947336

\bibitem{MR933023}
P.~P. Kulish and V.~D. Lipovsky, ``Hamiltonian structure of the
  {D}avey-{S}tewartson equation and {P}oisson brackets of scattering data,''
  \href{http://dx.doi.org/10.1016/0375-9601(88)90206-X}{{\em Phys. Lett. A}
  {\bfseries 127} no.~8-9, (1988) 413--417}.
  \url{http://dx.doi.org/10.1016/0375-9601(88)90206-X}. MR933023

\bibitem{MR976813}
P.~P. Kulish, ``The quantum superalgebra {${\rm osp}(2|1)$},''
  \href{http://dx.doi.org/10.1007/BF01101123}{{\em Zap. Nauchn. Sem. Leningrad.
  Otdel. Mat. Inst. Steklov. (LOMI)} {\bfseries 169} no.~Voprosy Kvant. Teor.
  Polya i Statist. Fiz. 8, (1988) 95--106, 188}.
  \url{http://dx.doi.org/10.1007/BF01101123}. MR976813

\bibitem{MR1010993}
P.~P. Kulish and N.~Y. Reshetikhin, ``Universal {$R$}-matrix of the quantum
  superalgebra {${\rm osp}(2|1)$},''
  \href{http://dx.doi.org/10.1007/BF00401868}{{\em Lett. Math. Phys.}
  {\bfseries 18} no.~2, (1989) 143--149}.
  \url{http://dx.doi.org/10.1007/BF00401868}. MR1010993

\bibitem{MR1091758}
P.~P. Kulish, ``Quantum {L}ie superalgebras and supergroups,'' in {\em Problems
  of modern quantum field theory ({A}lushta, 1989)}, Res. Rep. Phys.,
  pp.~14--21.
\newblock Springer, Berlin, 1989. MR1091758

\bibitem{MR1036959}
M.~Chaichian and P.~Kulish, ``Quantum {L}ie superalgebras and
  {$q$}-oscillators,''
  \href{http://dx.doi.org/10.1016/0370-2693(90)92004-3}{{\em Phys. Lett. B}
  {\bfseries 234} no.~1-2, (1990) 72--80}.
  \url{http://dx.doi.org/10.1016/0370-2693(90)92004-3}. MR1036959

\bibitem{MR1048775}
P.~P. Kulish and E.~V. Damaskinsky, ``On the {$q$} oscillator and the quantum
  algebra {${\rm su}_q(1,1)$},'' {\em J. Phys. A} {\bfseries 23} no.~9, (1990)
  L415--L419. \url{http://stacks.iop.org/0305-4470/23/L415}. MR1048775

\bibitem{MR1063456}
M.~Chaichian, P.~Kulish, and J.~Lukierski, ``{$q$}-deformed {J}acobi identity,
  {$q$}-oscillators and {$q$}-deformed infinite-dimensional algebras,''
  \href{http://dx.doi.org/10.1016/0370-2693(90)91196-I}{{\em Phys. Lett. B}
  {\bfseries 237} no.~3-4, (1990) 401--406}.
  \url{http://dx.doi.org/10.1016/0370-2693(90)91196-I}. MR1063456

\bibitem{MR1067335}
P.~P. Kulish, ``Clebsch-{G}ordan coefficients for a quantum superalgebra of
  rank one. {I},'' \href{http://dx.doi.org/10.1007/BF01249334}{{\em Zap.
  Nauchn. Sem. Leningrad. Otdel. Mat. Inst. Steklov. (LOMI)} {\bfseries 180}
  no.~Voprosy Kvant. Teor. Polya i Statist. Fiz. 9, (1990) 76--88, 180}.
  \url{http://dx.doi.org/10.1007/BF01249334}. MR1067335

\bibitem{MR1067336}
P.~P. Kulish, ``A two-parameter quantum group and a gauge transformation,''
  \href{http://dx.doi.org/10.1007/BF01249335}{{\em Zap. Nauchn. Sem. Leningrad.
  Otdel. Mat. Inst. Steklov. (LOMI)} {\bfseries 180} no.~Voprosy Kvant. Teor.
  Polya i Statist. Fiz. 9, (1990) 89--93, 180}.
  \url{http://dx.doi.org/10.1007/BF01249335}. MR1067336

\bibitem{MR1067477}
M.~Chaichian, D.~Ellinas, and P.~Kulish, ``Quantum algebra as the dynamical
  symmetry of the deformed {J}aynes-{C}ummings model,''
  \href{http://dx.doi.org/10.1103/PhysRevLett.65.980}{{\em Phys. Rev. Lett.}
  {\bfseries 65} no.~8, (1990) 980--983}.
  \url{http://dx.doi.org/10.1103/PhysRevLett.65.980}. MR1067477

\bibitem{MR1076342}
M.~Boiti, P.~Kulish, and F.~Pempinelli, ``Scattering of wall solitons in
  {$2+1$} dimensions,''
  \href{http://dx.doi.org/10.1016/0167-2789(90)90161-H}{{\em Phys. D}
  {\bfseries 44} no.~3, (1990) 557--564}.
  \url{http://dx.doi.org/10.1016/0167-2789(90)90161-H}. MR1076342

\bibitem{MR1106831}
P.~P. Kulish, ``Contraction of quantum algebras, and {$q$}-oscillators,''
  \href{http://dx.doi.org/10.1007/BF01018504}{{\em Teoret. Mat. Fiz.}
  {\bfseries 86} no.~1, (1991) 157--160}.
  \url{http://dx.doi.org/10.1007/BF01018504}. MR1106831

\bibitem{MR1111675}
E.~V. Damaskinsky and P.~P. Kulish, ``Deformed oscillators and their
  applications,'' \href{http://dx.doi.org/10.1007/BF01097496}{{\em Zap. Nauchn.
  Sem. Leningrad. Otdel. Mat. Inst. Steklov. (LOMI)} {\bfseries 189}
  no.~Voprosy Kvant. Teor. Polya i Statist. Fiz. 10, (1991) 37--74, 183}.
  \url{http://dx.doi.org/10.1007/BF01097496}. MR1111675

\bibitem{MR1114273}
M.~Chaichian, P.~Kulish, and J.~Lukierski, ``Supercovariant systems of
  {$q$}-oscillators and {$q$}-supersymmetric {H}amiltonians,''
  \href{http://dx.doi.org/10.1016/0370-2693(91)90640-C}{{\em Phys. Lett. B}
  {\bfseries 262} no.~1, (1991) 43--48}.
  \url{http://dx.doi.org/10.1016/0370-2693(91)90640-C}. MR1114273

\bibitem{MR1117592}
P.~P. Kulish and E.~K. Sklyanin, ``The general {$U_q[{\rm sl}(2)]$} invariant
  {$XXZ$} integrable quantum spin chain,'' {\em J. Phys. A} {\bfseries 24}
  no.~8, (1991) L435--L439. 
  \url{http://stacks.iop.org/0305-4470/24/L435}. MR1117592

\bibitem{MR1140211}
P.~P. Kulish, \href{http://dx.doi.org/10.1007/3-540-54040-7_106}{``Quantum
  algebras and symmetries of dynamical systems,''} in {\em Group theoretical
  methods in physics ({M}oscow, 1990)}, vol.~382 of {\em Lecture Notes in
  Phys.}, pp.~195--198.
\newblock Springer, Berlin, 1991.
\newblock \url{http://dx.doi.org/10.1007/3-540-54040-7_106}. MR1140211

\bibitem{MR1142025}
E.~Celeghini, R.~Giachetti, P.~P. Kulish, E.~Sorace, and M.~Tarlini, ``Hopf
  superalgebra contractions and {$R$}-matrix for fermions,'' {\em J. Phys. A}
  {\bfseries 24} no.~24, (1991) 5675--5682.
  \url{http://stacks.iop.org/0305-4470/24/5675}. MR1142025

\bibitem{MR1146012}
M.~Chaichian, P.~Kulish, and J.~Lukierski, ``Supercovariant
  {$q$}-oscillators,'' in {\em Nonlinear fields: classical, random,
  semiclassical ({K}arpacz, 1991)}, pp.~336--345.
\newblock World Sci. Publ., River Edge, NJ, 1991. MR1146012

\bibitem{MR1146146}
P.~P. Kulish, ``Finite-dimensional {Z}amolodchikov-{F}addeev algebra and
  {$q$}-oscillators,''
  \href{http://dx.doi.org/10.1016/0375-9601(91)90543-H}{{\em Phys. Lett. A}
  {\bfseries 161} no.~1, (1991) 50--52}.
  \url{http://dx.doi.org/10.1016/0375-9601(91)90543-H}. MR1146146

\bibitem{MR1191204}
M.~Chaichian and P.~Kulish, ``Quantum superalgebras, {$q$}-oscillators and
  applications,'' in {\em Nonperturbative methods in low-dimensional quantum
  field theories ({D}ebrecen, 1990)}, pp.~213--236.
\newblock World Sci. Publ., River Edge, NJ, 1991. MR1191204

\bibitem{MR1168674}
E.~V. Damaskinsky and P.~P. Kulish, ``Hermite {$q$}-polynomials and
  {$q$}-oscillators,'' \href{http://dx.doi.org/10.1007/BF02367234}{{\em Zap.
  Nauchn. Sem. S.-Peterburg. Otdel. Mat. Inst. Steklov. (POMI)} {\bfseries 199}
  no.~Voprosy Kvant. Teor. Polya Statist. Fiz. 11, (1992) 81--90, 185--186}.
  \url{http://dx.doi.org/10.1007/BF02367234}. MR1168674
  
\bibitem{MR1191263}
K.~Hikami, P.~P. Kulish, and M.~Wadati, ``Construction of integrable spin
  systems with long-range interactions,''
  \href{http://dx.doi.org/10.1143/JPSJ.61.3071}{{\em J. Phys. Soc. Japan}
  {\bfseries 61} no.~9, (1992) 3071--3076}.
  \url{http://dx.doi.org/10.1143/JPSJ.61.3071}. MR1191263

\bibitem{MR1193836}
P.~P. Kulish and E.~K. Sklyanin, ``Algebraic structures related to reflection
  equations,'' {\em J. Phys. A} {\bfseries 25} no.~22, (1992) 5963--5975.
  \url{http://stacks.iop.org/0305-4470/25/5963}. MR1193836

\bibitem{MR1239669}
P.~P. Kulish, ``Quantum algebras, {$q$}-deformed oscillators and related
  topics,'' in {\em Quantum group and quantum integrable systems}, Nankai
  Lectures Math. Phys., pp.~99--131.
\newblock World Sci. Publ., River Edge, NJ, 1992. MR1239669

\bibitem{MR1295926}
K.~Hikami, P.~P. Kulish, and M.~Wadati, ``Integrable spin systems with
  long-range interactions,''
  \href{http://dx.doi.org/10.1016/0960-0779(92)90029-M}{{\em Chaos Solitons
  Fractals} {\bfseries 2} no.~5, (1992) 543--550}.
  \url{http://dx.doi.org/10.1016/0960-0779(92)90029-M}. MR1295926

\bibitem{more19}
P.~P. Kulish, ``Deformed oscillators, non-commutative geometry, {Y}ang-{B}axter
  and reflection equations.''. Proceedings of XXI-DGMTP Conference, Tianjin,
  1992, 10 pp.

\bibitem{ISI:A1992BX67H00002}
P.~P. Kulish, ``Quantum groups - proceedings of workshops held in the
  {E}uler-{I}nternational-{M}athematical-{I}nstitute, {L}eningrad, {F}all 1990
  - preface,'' in {\em Quantum Groups}, {Kulish, P. P.}, ed., vol.~{1510} of
  {\em {Lecture notes in mathematics}}, p.~{R7}, Euler Int Math Inst.
\newblock {1992}. ISI:A1992BX67H00002

\bibitem{MR1198638}
P.~P. Kulish, R.~Sasaki, and C.~Schwiebert, ``Constant solutions of reflection
  equations and quantum groups,''
  \href{http://dx.doi.org/10.1063/1.530382}{{\em J. Math. Phys.} {\bfseries 34}
  no.~1, (1993) 286--304}. \url{http://dx.doi.org/10.1063/1.530382}. MR1198638

\bibitem{MR1208109}
P.~P. Kulish, ``On recent progress in quantum groups: an introductory review,''
  in {\em Jahrbuch \"Uberblicke {M}athematik, 1993}, pp.~97--124.
\newblock Friedr. Vieweg, Braunschweig, 1993. MR1208109

\bibitem{MR1215910}
P.~P. Kulish and R.~Sasaki, ``Covariance properties of reflection equation
  algebras,'' \href{http://dx.doi.org/10.1143/PTP.89.741}{{\em Progr. Theoret.
  Phys.} {\bfseries 89} no.~3, (1993) 741--761}.
  \url{http://dx.doi.org/10.1143/PTP.89.741}. MR1215910

\bibitem{MR1221730}
P.~P. Kulish, ``Quantum groups, {$Q$}-oscillators and covariant algebras,''
  \href{http://dx.doi.org/10.1007/BF01019325}{{\em Teoret. Mat. Fiz.}
  {\bfseries 94} no.~2, (1993) 193--199}.
  \url{http://dx.doi.org/10.1007/BF01019325}. MR1221730

\bibitem{MR1226066}
P.~P. Kulish, \href{http://dx.doi.org/10.1016/0083-6656(93)90009-9}{``Quantum
  groups and dynamical systems,''} in {\em Proceedings of the {I}nternational
  {S}ymposium on {Q}uantum {P}hysics and the {U}niverse ({T}okyo, 1992)},
  vol.~37, pp.~67--76.
\newblock 1993.
\newblock \url{http://dx.doi.org/10.1016/0083-6656(93)90009-9}. MR1226066

\bibitem{MR1230818}
M.~Chaichian, J.~F. Gomes, and P.~Kulish, ``Operator formulation of
  {$q$}-deformed dual string model,''
  \href{http://dx.doi.org/10.1016/0370-2693(93)90539-T}{{\em Phys. Lett. B}
  {\bfseries 311} no.~1-4, (1993) 93--97}.
  \url{http://dx.doi.org/10.1016/0370-2693(93)90539-T}. MR1230818

\bibitem{MR1255305}
P.~P. Kulish, ``Covariant noncommutative differential geometry,''
  \href{http://dx.doi.org/10.1007/BF02362779}{{\em Zap. Nauchn. Sem.
  S.-Peterburg. Otdel. Mat. Inst. Steklov. (POMI)} {\bfseries 205}
  no.~Differentsial'naya Geom. Gruppy Li i Mekh. 13, (1993) 85--91, 20, 180}.
  \url{http://dx.doi.org/10.1007/BF02362779}. MR1255305

\bibitem{MR1276292}
P.~P. Kulish, ``Reflection equation algebras and quantum groups,'' in {\em
  Quantum and non-commutative analysis ({K}yoto, 1992)}, vol.~16 of {\em Math.
  Phys. Stud.}, pp.~207--220.
\newblock Kluwer Acad. Publ., Dordrecht, 1993. MR1276292

\bibitem{MR1339517}
P.~P. Kulish, ``Quantum groups and quantum algebras as symmetries of dynamical
  systems,'' in {\em Quantum symmetries ({C}lausthal, 1991)}, pp.~51--74.
\newblock World Sci. Publ., River Edge, NJ, 1993. MR1339517

\bibitem{MR1290288}
P.~P. Kulish, S.~Rauch-Wojciechowski, and A.~V. Tsiganov, ``Restricted flows of
  the {K}d{V} hierarchy and {$r$}-matrix formalism,''
  \href{http://dx.doi.org/10.1142/S0217732394001921}{{\em Modern Phys. Lett. A}
  {\bfseries 9} no.~22, (1994) 2063--2073}.
  \url{http://dx.doi.org/10.1142/S0217732394001921}. MR1290288

\bibitem{MR1290824}
P.~P. Kulish, ``Representations of {$q$}-{M}inkowski space algebra,'' {\em
  Algebra i Analiz} {\bfseries 6} no.~2, (1994) 195--205. MR1290824

\bibitem{MR1299034}
J.~A. de~Azc\'arraga, P.~P. Kulish, and F.~R\'odenas, ``Reflection equations
  and {$q$}-{M}inkowski space algebras,''
  \href{http://dx.doi.org/10.1007/BF00750660}{{\em Lett. Math. Phys.}
  {\bfseries 32} no.~3, (1994) 173--182}.
  \url{http://dx.doi.org/10.1007/BF00750660}. MR1299034

\bibitem{MR1313739}
J.~A. de~Azc\'arraga, P.~P. Kulish, and F.~R\'odenas, ``Non-commutative
  geometry and covariance: from the quantum plane to quantum tensors,''
  \href{http://dx.doi.org/10.1007/BF01690450}{{\em Czechoslovak J. Phys.}
  {\bfseries 44} no.~11-12, (1994) 981--991}.
  \url{http://dx.doi.org/10.1007/BF01690450}. Quantum groups and physics
  (Prague, 1994). MR1313739

\bibitem{more20}
J.~A. de~Azc\'arraga, P.~P. Kulish, and F.~R\'odenas, ``Non-commutative
  geometry and covariance: from the quantum plane to quantum tensors.''.
  Preprint FTUV-94-54, Valencia, July, 1994, 5pp; Proceed. 20-th Yamada Intern.
  Colloquiium on Group Theory Methods in Physics, Kyoto, July, 1994.

\bibitem{more21}
P.~P. Kulish, ``Quantum groups and quantum homogeneous spaces.''. Proceedings
  of XXI-International Symposium on Elementary Particle Theory, DESY,
  Wendish-Rietz, 1993, 8pp; Preprint DESY 94-053 (1994).

\bibitem{MR1335177}
J.~A. de~Azc\'arraga, P.~P. Kulish, and F.~R\'odenas, ``On the physical
  contents of {$q$}-deformed {M}inkowski spaces,''
  \href{http://dx.doi.org/10.1016/0370-2693(95)00359-S}{{\em Phys. Lett. B}
  {\bfseries 351} no.~1-3, (1995) 123--130}.
  \url{http://dx.doi.org/10.1016/0370-2693(95)00359-S}. MR1335177

\bibitem{MR1349334}
Y.~S. Osipov, A.~A. Gonchar, S.~P. Novikov, V.~I. Arnol'd, G.~I. Marchuk, P.~P.
  Kulish, V.~S. Vladimirov, and E.~F. Mishchenko, ``Lyudvig {D}mitrievich
  {F}addeev (on the occasion of his sixtieth birthday),''
  \href{http://dx.doi.org/10.1070/RM1995v050n03ABEH002577}{{\em Uspekhi Mat.
  Nauk} {\bfseries 50} no.~3(303), (1995) 171--186}.
  \url{http://dx.doi.org/10.1070/RM1995v050n03ABEH002577}. MR1349334

\bibitem{MR1364848}
E.~V. Damaskinsky, P.~P. Kulish, and M.~A. Sokolov, ``Gauss decompositions of
  quantum groups and supergroups,''
  \href{http://dx.doi.org/10.1007/BF02364982}{{\em Zap. Nauchn. Sem.
  S.-Peterburg. Otdel. Mat. Inst. Steklov. (POMI)} {\bfseries 224} no.~Voprosy
  Kvant. Teor. Polya i Statist. Fiz. 13, (1995) 155--177, 338}.
  \url{http://dx.doi.org/10.1007/BF02364982}. MR1364848

\bibitem{MR1413741}
J.~A. de~Azc\'arraga, P.~P. Kulish, and F.~Rodenas, ``Dirac-like formulation of
  {$q$}-{L}orentz groups,'' in {\em Group theoretical methods in physics
  ({T}oyonaka, 1994)}, pp.~133--137.
\newblock World Sci. Publ., River Edge, NJ, 1995. MR1413741

\bibitem{MR1390845}
J.~A. de~Azc\'arraga, P.~P. Kulish, and F.~Rodenas, ``Quantum groups and
  deformed special relativity,''
  \href{http://dx.doi.org/10.1002/prop.2190440102}{{\em Fortschr. Phys.}
  {\bfseries 44} no.~1, (1996) 1--40}.
  \url{http://dx.doi.org/10.1002/prop.2190440102}. MR1390845

\bibitem{MR1401235}
P.~P. Kulish, S.~Rauch-Wojciechowski, and A.~V. Tsiganov, ``Stationary problems
  for equation of the {K}d{V} type and dynamical {$r$}-matrices,''
  \href{http://dx.doi.org/10.1063/1.531575}{{\em J. Math. Phys.} {\bfseries 37}
  no.~7, (1996) 3463--3482}. 
  \url{http://dx.doi.org/10.1063/1.531575}. MR1401235

\bibitem{MR1415854}
P.~P. Kulish, ``Quantum group covariant algebras,'' in {\em Quantum groups and
  their applications in physics ({V}arenna, 1994)}, vol.~127 of {\em Proc.
  Internat. School Phys. Enrico Fermi}, pp.~203--214.
\newblock IOS, Amsterdam, 1996. MR1415854

\bibitem{MR1428436}
M.~Chaichian and P.~P. Kulish,
  \href{http://dx.doi.org/10.1142/9789812830425_0006}{``Quantum group covariant
  systems,''} in {\em From field theory to quantum groups}, pp.~99--111.
\newblock World Sci. Publ., River Edge, NJ, 1996.
\newblock \url{http://dx.doi.org/10.1142/9789812830425_0006}. MR1428436

\bibitem{MR1477944}
P.~P. Kulish, \href{http://dx.doi.org/10.1007/BFb0102555}{``Yang-{B}axter
  equation and reflection equations in integrable models,''} in {\em
  Low-dimensional models in statistical physics and quantum field theory
  ({S}chladming, 1995)}, vol.~469 of {\em Lecture Notes in Phys.},
  pp.~125--144.
\newblock Springer, Berlin, 1996.
\newblock \url{http://dx.doi.org/10.1007/BFb0102555}. MR1477944

\bibitem{MR1684785}
E.~V. Damaskinsky, P.~P. Kulish, V.~D. Lyakhovsky, and M.~A. Sokolov, ``Gauss
  decomposition for quantum groups and duality,'' in {\em Symmetry methods in
  physics, {V}ol. 1 ({D}ubna, 1995)}, pp.~120--129.
\newblock Joint Inst. Nuclear Res., Dubna, 1996. MR1684785

\bibitem{more22}
P.~P. Kulish, ``Comment on q-oscillator coherent states.''. Preprint KTH-96-31,
  Stockholm, 1996, 4pp.

\bibitem{more23}
P.~P. Kulish, ``Quantum integrable systems.''. Lecture course, Royal Institute
  of Technology, Physics Department, Stockholm, 1996.

\bibitem{MR1446986}
E.~V. Damaskinsky and P.~P. Kulish, ``Irreducible representations of deformed
  oscillator algebra and {$q$}-special functions,''
  \href{http://dx.doi.org/10.1142/S0217751X97000207}{{\em Internat. J. Modern
  Phys. A} {\bfseries 12} no.~1, (1997) 153--158}.
  \url{http://dx.doi.org/10.1142/S0217751X97000207}. IWCQIS 96 
  (Dubna, 1996). MR1446986

\bibitem{MR1449302}
A.~P. Isaev and P.~P. Kulish, ``Tetrahedron reflection equations,''
  \href{http://dx.doi.org/10.1142/S0217732397000443}{{\em Modern Phys. Lett. A}
  {\bfseries 12} no.~6, (1997) 427--437}.
  \url{http://dx.doi.org/10.1142/S0217732397000443}. MR1449302

\bibitem{MR1456489}
A.~Stolin and P.~P. Kulish, ``New rational solutions of {Y}ang-{B}axter
  equation and deformed {Y}angians,''
  \href{http://dx.doi.org/10.1023/A:1021460515598}{{\em Czechoslovak J. Phys.}
  {\bfseries 47} no.~1, (1997) 123--129}.
  \url{http://dx.doi.org/10.1023/A:1021460515598}. Quantum groups and
  integrable systems, II (Prague, 1996). MR1456489
 
  
\bibitem{MR1474459}
M.~Chaichian, A.~P. Demichev, and P.~P. Kulish, ``Quasi-classical limit in
  {$q$}-deformed systems, non-commutativity and the {$q$}-path integral,''
  \href{http://dx.doi.org/10.1016/S0375-9601(97)00513-6}{{\em Phys. Lett. A}
  {\bfseries 233} no.~4-6, (1997) 251--260}.
  \url{http://dx.doi.org/10.1016/S0375-9601(97)00513-6}. MR1474459

\bibitem{MR1483888}
J.~A. de~Azc\'arraga, P.~P. Kulish, and F.~Rodenas, ``Twisted {$h$}-spacetimes
  and invariant equations,''
  \href{http://dx.doi.org/10.1007/s002880050579}{{\em Z. Phys. C} {\bfseries
  76} no.~3, (1997) 567--576}. 
  \url{http://dx.doi.org/10.1007/s002880050579}. MR1483888

\bibitem{MR1608809}
P.~P. Kulish and A.~A. Stolin, ``Deformed {Y}angians and integrable models,''
  \href{http://dx.doi.org/10.1023/A:1022869414679}{{\em Czechoslovak J. Phys.}
  {\bfseries 47} no.~12, (1997) 1207--1212}.
  \url{http://dx.doi.org/10.1023/A:1022869414679}. Quantum groups and
  integrable systems, II (Prague, 1997). MR1608809

\bibitem{MR1627837}
P.~N. Bibikov and P.~P. Kulish, ``Dirac operators on the quantum group {${\rm
  SU}_q(2)$} and the quantum sphere,''
  \href{http://dx.doi.org/10.1007/BF02675726}{{\em Zap. Nauchn. Sem.
  S.-Peterburg. Otdel. Mat. Inst. Steklov. (POMI)} {\bfseries 245} no.~Vopr.
  Kvant. Teor. Polya i Stat. Fiz. 14, (1997) 49--65, 283}.
  \url{http://dx.doi.org/10.1007/BF02675726}. MR1627837

\bibitem{more24}
P.~P. Kulish, ``Quantum groups and their representations.''. Lecture course,
  Department of Mathematics, Royal Institute of Technology, Stockholm, 1997.

\bibitem{MR1627540}
E.~Celeghini and P.~P. Kulish, ``Twist deformation of the rank-one {L}ie
  superalgebra,'' \href{http://dx.doi.org/10.1088/0305-4470/31/4/001}{{\em J.
  Phys. A} {\bfseries 31} no.~4, (1998) L79--L84}.
  \url{http://dx.doi.org/10.1088/0305-4470/31/4/001}. MR1627540

\bibitem{MR1681992}
P.~P. Kulish and V.~D. Lyakhovsky, ``Classical and quantum duality in
  {J}ordanian quantizations,''
  \href{http://dx.doi.org/10.1023/A:1021613407866}{{\em Czechoslovak J. Phys.}
  {\bfseries 48} no.~11, (1998) 1415--1421}.
  \url{http://dx.doi.org/10.1023/A:1021613407866}. Quantum groups and
  integrable systems (Prague, 1998). MR1681992

\bibitem{MR1700692}
E.~V. Damaskinsky, P.~P. Kulish, and M.~Cha\u\i~chian, ``Dynamical systems
  associated with the {C}remmer-{G}ervais {$R$}-matrix,''
  \href{http://dx.doi.org/10.1007/BF02557124}{{\em Teoret. Mat. Fiz.}
  {\bfseries 116} no.~1, (1998) 101--112}.
  \url{http://dx.doi.org/10.1007/BF02557124}. MR1700692

\bibitem{MR1737872}
P.~P. Kulish and M.~A. Sokolov, ``Gauss decomposition of quantum groups and a
  model of discrete dynamics,''
  \href{http://dx.doi.org/10.1023/A:1011301007755}{{\em Zap. Nauchn. Sem.
  S.-Peterburg. Otdel. Mat. Inst. Steklov. (POMI)} {\bfseries 251} no.~Vopr.
  Kvant. Teor. Polya i Stat. Fiz. 15, (1998) 105--114, 275--276 (1999)}.
  \url{http://dx.doi.org/10.1023/A:1011301007755}. MR1737872

\bibitem{MR1765492}
P.~P. Kulish, ``Twisting of quantum groups and integrable models,'' in {\em
  Particles, fields, and gravitation (\L \'od\'z, 1998)}, vol.~453 of {\em AIP
  Conf. Proc.}, pp.~75--85.
\newblock Amer. Inst. Phys., Woodbury, NY, 1998. MR1765492

\bibitem{MR1682301}
P.~P. Kulish and A.~I. Mudrov, ``Universal {$R$}-matrix for esoteric quantum
  groups,'' \href{http://dx.doi.org/10.1023/A:1007538903995}{{\em Lett. Math.
  Phys.} {\bfseries 47} no.~2, (1999) 139--148}.
  \url{http://dx.doi.org/10.1023/A:1007538903995}. MR1682301

\bibitem{MR1708368}
P.~P. Kulish, V.~D. Lyakhovsky, and A.~I. Mudrov, ``Extended {J}ordanian twists
  for {L}ie algebras,'' \href{http://dx.doi.org/10.1063/1.532987}{{\em J. Math.
  Phys.} {\bfseries 40} no.~9, (1999) 4569--4586}.
  \url{http://dx.doi.org/10.1063/1.532987}. MR1708368

\bibitem{MR1719483}
P.~P. Kulish, \href{http://dx.doi.org/10.1142/S0217979299002770}{``Symmetries
  related to {Y}ang-{B}axter equation and reflection equations,''} in {\em
  Proceedings of the {E}uroconference on {N}ew {S}ymmetries in {S}tatistical
  {M}echanics and {C}ondensed {M}atter {P}hysics ({T}orino, 1998)}, vol.~13,
  pp.~2943--2951.
\newblock 1999.
\newblock \url{http://dx.doi.org/10.1142/S0217979299002770}. MR1719483

\bibitem{MR1726595}
P.~P. Kulish, ``Deformed oscillator and path integrals,'' in {\em Path
  integrals from pe{V} to {T}e{V} ({F}lorence, 1998)}, pp.~217--220.
\newblock World Sci. Publ., River Edge, NJ, 1999. MR1726595

\bibitem{MR1732615}
P.~P. Kulish, V.~D. Lyakhovsky, and M.~A. del Olmo, ``Chains of twists for
  classical {L}ie algebras,''
  \href{http://dx.doi.org/10.1088/0305-4470/32/49/308}{{\em J. Phys. A}
  {\bfseries 32} no.~49, (1999) 8671--8684}.
  \url{http://dx.doi.org/10.1088/0305-4470/32/49/308}. MR1732615

\bibitem{MR1782555}
P.~P. Kulish and A.~I. Mudrov, ``Twist-like geometries on a quantum {M}inkowski
  space,'' {\em Tr. Mat. Inst. Steklova} {\bfseries 226} no.~Mat. Fiz. Probl.
  Kvantovo\u\i \ Teor. Polya, (1999) 97--111. MR1782555

\bibitem{more25}
P.~P. Kulish, ``Super-{J}ordanian deformation of the orthosymplectic {L}ie
  superalgebras.''. Preprint DFF-1998/11, 10p.; Mod. Phys. Lett. (1999) ;
  math.QA/9806104.

\bibitem{MR1767500}
V.~V. Borzov, E.~V. Damaskinsky, and P.~P. Kulish, ``Construction of the
  spectral measure for deformed oscillator position operator in the case of
  undetermined {H}amburger moment problem,''
  \href{http://dx.doi.org/10.1142/S0129055X0000023X}{{\em Rev. Math. Phys.}
  {\bfseries 12} no.~5, (2000) 691--710}.
  \url{http://dx.doi.org/10.1142/S0129055X0000023X}. MR1767500
  
\bibitem{MR1772994}
P.~P. Kulish, ``Twisting of quantum groups and integrable systems,'' in {\em
  Proceedings of the {W}orkshop on {N}onlinearity, {I}ntegrability and {A}ll
  {T}hat: {T}wenty {Y}ears after {NEEDS} '79 ({G}allipoli, 1999)},
  pp.~304--310.
\newblock World Sci. Publ., River Edge, NJ, 2000. MR1772994

\bibitem{more26}
P.~P. Kulish, ``Twisting of quantum groups and integrable models.''. Proceed.
  Lodz Conference: Particles, Fields, and Gravitation; J. Rembielinski (ed),
  Amer. Inst. Physics, 1998, 75-85.

\bibitem{MR1774650}
P.~P. Kulish and A.~M. Nikitin, ``Invariants of {B}-type links via an extension
  of the {K}auffman bracket,''
  \href{http://dx.doi.org/10.1023/A:1021198210356}{{\em Zap. Nauchn. Sem.
  S.-Peterburg. Otdel. Mat. Inst. Steklov. (POMI)} {\bfseries 266} no.~Teor.
  Predst. Din. Sist. Komb. i Algoritm. Metody. 5, (2000) 107--130, 338}.
  \url{http://dx.doi.org/10.1023/A:1021198210356}. MR1774650

\bibitem{MR1784221}
P.~P. Kulish and V.~D. Lyakhovsky, ``Jordanian twists on deformed carrier
  subspaces,'' \href{http://dx.doi.org/10.1088/0305-4470/33/31/103}{{\em J.
  Phys. A} {\bfseries 33} no.~31, (2000) L279--L285}.
  \url{http://dx.doi.org/10.1088/0305-4470/33/31/103}. MR1784221

\bibitem{MR1798773}
P.~P. Kulish and A.~I. Mudrov, ``On twisting solutions to the {Y}ang-{B}axter
  equation,'' \href{http://dx.doi.org/10.1023/A:1022885317520}{{\em
  Czechoslovak J. Phys.} {\bfseries 50} no.~1, (2000) 115--122}.
  \url{http://dx.doi.org/10.1023/A:1022885317520}. Quantum groups and
  integrable systems (Prague, 1999). MR1798773

\bibitem{MR1805860}
E.~V. Damaskinsky and P.~P. Kulish, ``Symmetries associated with the
  {Y}ang-{B}axter equation and the reflection equation,''
  \href{http://dx.doi.org/10.1023/A:1022699729324}{{\em Zap. Nauchn. Sem.
  S.-Peterburg. Otdel. Mat. Inst. Steklov. (POMI)} {\bfseries 269} no.~Vopr.
  Kvant. Teor. Polya i Stat. Fiz. 16, (2000) 180--192, 368--369}.
  \url{http://dx.doi.org/10.1023/A:1022699729324}. MR1805860

\bibitem{MR1805861}
E.~V. Damaskinsky, P.~P. Kulish, and M.~A. Sokolov, ``On the structure of
  co-boundary {$R$}-matrices of classical series,''
  \href{http://dx.doi.org/10.1023/A:1022651813395}{{\em Zap. Nauchn. Sem.
  S.-Peterburg. Otdel. Mat. Inst. Steklov. (POMI)} {\bfseries 269} no.~Vopr.
  Kvant. Teor. Polya i Stat. Fiz. 16, (2000) 193--206, 369}.
  \url{http://dx.doi.org/10.1023/A:1022651813395}. MR1805861

\bibitem{MR1806276}
P.~P. Kulish, V.~D. Lyakhovsky, and A.~Stolin, ``Full chains of twists for
  orthogonal algebras,'' \href{http://dx.doi.org/10.1023/A:1022873326934}{{\em
  Czechoslovak J. Phys.} {\bfseries 50} no.~11, (2000) 1291--1296}.
  \url{http://dx.doi.org/10.1023/A:1022873326934}. Quantum groups and
  integrable systems (Prague, 2000). MR1806276

\bibitem{MR1845802}
P.~P. Kulish and N.~Manojlovi\'c, ``Bethe vectors of the {${\rm osp}(1|2)$}
  {G}audin model,'' \href{http://dx.doi.org/10.1023/A:1010950003268}{{\em Lett.
  Math. Phys.} {\bfseries 55} no.~1, (2001) 77--95}.
  \url{http://dx.doi.org/10.1023/A:1010950003268}. MR1845802

\bibitem{MR1855095}
P.~P. Kulish and N.~Manojlovi\'c, ``Creation operators and {B}ethe vectors of
  the {$\rm osp(1|2)$} {G}audin model,''
  \href{http://dx.doi.org/10.1063/1.1398584}{{\em J. Math. Phys.} {\bfseries
  42} no.~10, (2001) 4757--4778}. 
  \url{http://dx.doi.org/10.1063/1.1398584}. MR1855095

\bibitem{MR1855355}
V.~D. Lyakhovsky, A.~Stolin, and P.~P. Kulish, ``Chains of {F}robenius
  subalgebras of {${\rm so}(M)$} and the corresponding twists,''
  \href{http://dx.doi.org/10.1063/1.1402177}{{\em J. Math. Phys.} {\bfseries
  42} no.~10, (2001) 5006--5019}. 
  \url{http://dx.doi.org/10.1063/1.1402177}. MR1855355

\bibitem{MR1894076}
P.~P. Kulish and A.~I. Mudrov, ``Quantization of inhomogeneous {L}ie
  bialgebras,'' \href{http://dx.doi.org/10.1016/S0393-0440(01)00073-0}{{\em J.
  Geom. Phys.} {\bfseries 42} no.~1-2, (2002) 64--77}.
  \url{http://dx.doi.org/10.1016/S0393-0440(01)00073-0}. MR1894076

\bibitem{MR1921987}
D.~N. Ananikyan, P.~P. Kulish, and V.~D. Lyakhovsky, ``Full chains of twists
  for symplectic algebras,'' {\em Algebra i Analiz} {\bfseries 14} no.~3,
  (2002) 27--54. MR1921987

\bibitem{MR1948815}
E.~V. Damaskinsky, P.~P. Kulish, and M.~A. Sokolov, ``Unified quantization of
  three-dimensional bialgebras,''
  \href{http://dx.doi.org/10.1023/B:JOTH.0000049571.26334.b7}{{\em Zap. Nauchn.
  Sem. S.-Peterburg. Otdel. Mat. Inst. Steklov. (POMI)} {\bfseries 291}
  no.~Vopr. Kvant. Teor. Polya i Stat. Fiz. 17, (2002) 169--184, 281}.
  \url{http://dx.doi.org/10.1023/B:JOTH.0000049571.26334.b7}. MR1948815

\bibitem{MR1948816}
P.~P. Kulish and A.~M. Zeitlin, ``Group-theoretic structure and the inverse
  scattering problem method for the super {K}d{V} equation,''
  \href{http://dx.doi.org/10.1023/B:JOTH.0000049572.41993.9f}{{\em Zap. Nauchn.
  Sem. S.-Peterburg. Otdel. Mat. Inst. Steklov. (POMI)} {\bfseries 291}
  no.~Vopr. Kvant. Teor. Polya i Stat. Fiz. 17, (2002) 185--205, 281}.
  \url{http://dx.doi.org/10.1023/B:JOTH.0000049572.41993.9f}. MR1948816

\bibitem{MR1948818}
A.~A. Stolin, P.~P. Kulish, and E.~V. Damaskinsky, ``On the construction of the
  universal twist element,''
  \href{http://dx.doi.org/10.1023/B:JOTH.0000049574.67818.ed}{{\em Zap. Nauchn.
  Sem. S.-Peterburg. Otdel. Mat. Inst. Steklov. (POMI)} {\bfseries 291}
  no.~Vopr. Kvant. Teor. Polya i Stat. Fiz. 17, (2002) 228--244, 282}.
  \url{http://dx.doi.org/10.1023/B:JOTH.0000049574.67818.ed}. MR1948818

\bibitem{MR1964909}
M.~Chaichian and P.~P. Kulish,
  \href{http://dx.doi.org/10.1142/9789812777065_0024}{``Spin {H}amiltonians,
  quantum groups and reaction-diffusion processes,''} in {\em Multiple facets
  of quantization and supersymmetry}, pp.~310--319.
\newblock World Sci. Publ., River Edge, NJ, 2002.
\newblock \url{http://dx.doi.org/10.1142/9789812777065_0024}. MR1964909

\bibitem{MR1966936}
P.~P. Kulish and A.~Stolin, ``Twists in {H}opf algebras and {$RT_1T_2=T_2T_1$}
  relations,'' \href{http://dx.doi.org/10.1023/A:1021341121820}{{\em
  Czechoslovak J. Phys.} {\bfseries 52} no.~11, (2002) 1255--1260}.
  \url{http://dx.doi.org/10.1023/A:1021341121820}. Quantum groups and
  integrable systems (Prague, 2002). MR1966936

\bibitem{more27}
A.~Stolin, P.~P. Kulish, and E.~V. Damaskinsky, ``On reconstruction of
  universal twist from ${R}$-matrix.''. Preprint of St. Petersburg Steklov
  Math. Inst. (2002); Zap. Nauch. Semin. PDMI 291 (2002) 228 - 244;
  math.QA/0307306.

\bibitem{MR1952209}
P.~P. Kulish and N.~Manojlovi\'c, ``Trigonometric {${\rm osp}(1|2)$} {G}audin
  model,'' \href{http://dx.doi.org/10.1063/1.1531250}{{\em J. Math. Phys.}
  {\bfseries 44} no.~2, (2003) 676--700}.
  \url{http://dx.doi.org/10.1063/1.1531250}. MR1952209

\bibitem{MR1992884}
J.~Donin, P.~P. Kulish, and A.~I. Mudrov, ``On a universal solution to the
  reflection equation,'' \href{http://dx.doi.org/10.1023/A:1024438101617}{{\em
  Lett. Math. Phys.} {\bfseries 63} no.~3, (2003) 179--194}.
  \url{http://dx.doi.org/10.1023/A:1024438101617}. MR1992884

\bibitem{MR2006441}
P.~P. Kulish, ``On spin systems related to the {T}emperley-{L}ieb algebra,''
  \href{http://dx.doi.org/10.1088/0305-4470/36/38/101}{{\em J. Phys. A}
  {\bfseries 36} no.~38, (2003) L489--L493}.
  \url{http://dx.doi.org/10.1088/0305-4470/36/38/101}. MR2006441

\bibitem{MR2021097}
P.~P. Kulish, ``Quantum groups and integrable models,'' in {\em Factorization
  and integrable systems ({F}aro, 2000)}, vol.~141 of {\em Oper. Theory Adv.
  Appl.}, pp.~131--154.
\newblock Birkh\"auser, Basel, 2003. MR2021097

\bibitem{more28}
P.~P. Kulish, ``Introduction to classical and quantum integrable systems.''.

\bibitem{MR2065667}
E.~Celeghini and P.~P. Kulish, ``Deformation of orthosymplectic {L}ie
  superalgebra {${\rm osp}(1|4)$},''
  \href{http://dx.doi.org/10.1088/0305-4470/37/20/L01}{{\em J. Phys. A}
  {\bfseries 37} no.~20, (2004) L211--L216}.
  \url{http://dx.doi.org/10.1088/0305-4470/37/20/L01}. MR2065667

\bibitem{MR2093604}
P.~P. Kulish and A.~M. Zeitlin, ``Superconformal field theory and {SUSY}
  {$N=1$} {K}d{V} hierarchy. {I}. {V}ertex operators and {Y}ang-{B}axter
  equation,'' \href{http://dx.doi.org/10.1016/j.physletb.2004.07.019}{{\em
  Phys. Lett. B} {\bfseries 597} no.~2, (2004) 229--236}.
  \url{http://dx.doi.org/10.1016/j.physletb.2004.07.019}. MR2093604

\bibitem{MR2102324}
P.~P. Kulish and A.~M. Zeitlin, ``Integrable structure of superconformal field
  theory and quantum super-{K}d{V} theory,''
  \href{http://dx.doi.org/10.1016/j.physletb.2003.12.008}{{\em Phys. Lett. B}
  {\bfseries 581} no.~1-2, (2004) 125--132}.
  \url{http://dx.doi.org/10.1016/j.physletb.2003.12.008}. MR2102324

\bibitem{MR2123761}
M.~Chaichian, P.~P. Kulish, K.~Nishijima, and A.~Tureanu, ``On a
  {L}orentz-invariant interpretation of noncommutative space-time and its
  implications on noncommutative {QFT},''
  \href{http://dx.doi.org/10.1016/j.physletb.2004.10.045}{{\em Phys. Lett. B}
  {\bfseries 604} no.~1-2, (2004) 98--102}.
  \url{http://dx.doi.org/10.1016/j.physletb.2004.10.045}. MR2123761

\bibitem{ISI:000221185600049}
P.~P. Kulish and A.~M. Zeitlin, ``{Quantization of integrable models with
  hidden symmetries: super-KdV equation},''
  \href{http://dx.doi.org/{10.1080/09500340410001664106}}{{\em Journal of
  Modern Optics} {\bfseries {51}} no.~{6-7}, ({APR 15}, {2004}) {1107--1108}}.
  {5th Workshop on Mysteries, Puzzles and Paradoxes in Quantum Mechanics
  (MPPinQM-5), Palazzo Feltrinelli, Gargnano, ITALY, SEP 01-05, 
  2003}. ISI:000221185600049
  
\bibitem{MR2123217}
P.~P. Kulish and A.~M. Zeitlin, ``Superconformal field theory and {SUSY}
  {$N=1$} {K}d{V} hierarchy. {II}. {T}he {$Q$}-operator,''
  \href{http://dx.doi.org/10.1016/j.nuclphysb.2004.12.031}{{\em Nuclear Phys.
  B} {\bfseries 709} no.~3, (2005) 578--591}.
  \url{http://dx.doi.org/10.1016/j.nuclphysb.2004.12.031}. MR2123217

\bibitem{MR2141776}
P.~P. Kulish and A.~M. Zeitlin, ``The quantum inverse problem method and
  (super)conformal field theory,''
  \href{http://dx.doi.org/10.1007/s11232-005-0054-5}{{\em Teoret. Mat. Fiz.}
  {\bfseries 142} no.~2, (2005) 252--264}.
  \url{http://dx.doi.org/10.1007/s11232-005-0054-5}. MR2141776

\bibitem{MR2153657}
P.~P. Kulish and A.~M. Zeitlin, ``Quantum supersymmetric {T}oda-m{K}d{V}
  hierarchies,'' \href{http://dx.doi.org/10.1016/j.nuclphysb.2005.06.002}{{\em
  Nuclear Phys. B} {\bfseries 720} no.~3, (2005) 289--306}.
  \url{http://dx.doi.org/10.1016/j.nuclphysb.2005.06.002}. MR2153657

\bibitem{MR2160324}
P.~P. Kulish and P.~D. Ryasichenko, ``A spin chain connected with the quantum
  superalgebra {${\rm sl}_q(1|1)$},''
  \href{http://dx.doi.org/10.1007/s10958-006-0339-8}{{\em Zap. Nauchn. Sem.
  S.-Peterburg. Otdel. Mat. Inst. Steklov. (POMI)} {\bfseries 325} no.~Teor.
  Predst. Din. Sist. Komb. i Algoritm. Metody. 12, (2005) 146--162, 246--247}.
  \url{http://dx.doi.org/10.1007/s10958-006-0339-8}. MR2160324

\bibitem{MR2184025}
P.~P. Kulish, \href{http://dx.doi.org/10.1090/conm/391/07331}{``Noncommutative
  geometry and quantum field theory,''} in {\em Noncommutative geometry and
  representation theory in mathematical physics}, vol.~391 of {\em Contemp.
  Math.}, pp.~213--221.
\newblock Amer. Math. Soc., Providence, RI, 2005.
\newblock \url{http://dx.doi.org/10.1090/conm/391/07331}. MR2184025
 

\bibitem{MR2213317}
P.~P. Kulish and A.~I. Mudrov, ``Baxterization of solutions to reflection
  equation with {H}ecke {$R$}-matrix,''
  \href{http://dx.doi.org/10.1007/s11005-005-0043-5}{{\em Lett. Math. Phys.}
  {\bfseries 75} no.~2, (2006) 151--170}.
  \url{http://dx.doi.org/10.1007/s11005-005-0043-5}. MR2213317

\bibitem{MR2236645}
P.~P. Kulish, V.~D. Lyakhovsky, and M.~E. Samsonov, ``Twists in {$U(sl_3)$} and
  their quantizations,''
  \href{http://dx.doi.org/10.1088/0305-4470/39/24/005}{{\em J. Phys. A}
  {\bfseries 39} no.~24, (2006) 7669--7692}.
  \url{http://dx.doi.org/10.1088/0305-4470/39/24/005}. MR2236645

\bibitem{MR2269757}
P.~P. Kulish, ``Twist of quantum groups and noncommutative field theory,''
  \href{http://dx.doi.org/10.1007/s10958-007-0166-6}{{\em Zap. Nauchn. Sem.
  S.-Peterburg. Otdel. Mat. Inst. Steklov. (POMI)} {\bfseries 335} no.~Vopr.
  Kvant. Teor. Polya i Stat. Fiz. 19, (2006) 188--204}.
  \url{http://dx.doi.org/10.1007/s10958-007-0166-6}. MR2269757

\bibitem{MR2283730}
L.~Frappat, P.~Kulish, E.~Ragoucy, V.~Rivasseau, and P.~Sorba, ``Editorial: in
  memoriam {D}aniel {A}rnaudon,''
  \href{http://dx.doi.org/10.1007/s00023-006-0279-3}{{\em Ann. Henri
  Poincar\'e} {\bfseries 7} no.~7-8, (2006) 1213--1216}.
  \url{http://dx.doi.org/10.1007/s00023-006-0279-3}. MR2283730

\bibitem{MR2349627}
P.~P. Kulish and A.~I. Mudrov,
  \href{http://dx.doi.org/10.1090/conm/433/08331}{``Dynamical reflection
  equation,''} in {\em Quantum groups}, vol.~433 of {\em Contemp. Math.},
  pp.~281--309.
\newblock Amer. Math. Soc., Providence, RI, 2007.
\newblock \url{http://dx.doi.org/10.1090/conm/433/08331}. MR2349627

\bibitem{MR2371268}
A.~Fring, P.~Kulish, N.~Manojlovic, Z.~Nagy, J.~Nunes~da Costa, and
  H.~Samtleben, ``Special issue on recent developments in infinite dimensional
  algebras and their applications to quantum integrable systems,'' {\em J.
  Phys. A} {\bfseries 40} no.~33, (2007) i--ii. MR2371268

\bibitem{MR2458890}
P.~P. Kulish and N.~Mano\u\i~lovich, ``Quantum algebras with representation
  ring of type {${\rm sl}(2)$},''
  \href{http://dx.doi.org/10.1007/s10958-008-9011-9}{{\em Zap. Nauchn. Sem.
  S.-Peterburg. Otdel. Mat. Inst. Steklov. (POMI)} {\bfseries 347} no.~Vopr.
  Kvant. Teor. Polya i Stat. Fiz. 20, (2007) 167--177, 242}.
  \url{http://dx.doi.org/10.1007/s10958-008-9011-9}. MR2458890

\bibitem{MR2458891}
P.~P. Kulish and P.~D. Ryasichenko, ``The algebraic {B}ethe ansatz for a
  seven-vertex model,'' \href{http://dx.doi.org/10.1007/s10958-008-9012-8}{{\em
  Zap. Nauchn. Sem. S.-Peterburg. Otdel. Mat. Inst. Steklov. (POMI)} {\bfseries
  347} no.~Vopr. Kvant. Teor. Polya i Stat. Fiz. 20, (2007) 178--186, 242}.
  \url{http://dx.doi.org/10.1007/s10958-008-9012-8}. MR2458891

\bibitem{MR2392870}
P.~P. Kulish, N.~Manojlovic, and Z.~Nagy, ``Quantum symmetry algebras of spin
  systems related to {T}emperley-{L}ieb {$R$}-matrices,''
  \href{http://dx.doi.org/10.1063/1.2873025}{{\em J. Math. Phys.} {\bfseries
  49} no.~2, (2008) 023510, 9}. 
  \url{http://dx.doi.org/10.1063/1.2873025}. MR2392870

\bibitem{MR2412281}
M.~Chaichian, P.~P. Kulish, A.~Tureanu, R.~B. Zhang, and X.~Zhang,
  ``Noncommutative fields and actions of twisted {P}oincar\'e algebra,''
  \href{http://dx.doi.org/10.1063/1.2907580}{{\em J. Math. Phys.} {\bfseries
  49} no.~4, (2008) 042302, 16}. 
  \url{http://dx.doi.org/10.1063/1.2907580}. MR2412281
  
\bibitem{MR2435011}
P.~P. Kulish, ``Models solvable by {B}ethe ansatz,''
  \href{http://dx.doi.org/10.4303/jglta/S080317}{{\em J. Gen. Lie Theory Appl.}
  {\bfseries 2} no.~3, (2008) 190--200}.
  \url{http://dx.doi.org/10.4303/jglta/S080317}. MR2435011

\bibitem{MR2451509}
B.~Aneva, M.~Chaichian, and P.~P. Kulish, ``From quantum affine symmetry to the
  boundary {A}skey-{W}ilson algebra and the reflection equation,''
  \href{http://dx.doi.org/10.1088/1751-8113/41/13/135201}{{\em J. Phys. A}
  {\bfseries 41} no.~13, (2008) 135201, 18}.
  \url{http://dx.doi.org/10.1088/1751-8113/41/13/135201}. MR2451509

\bibitem{MR2452174}
A.~Fring, P.~P. Kulish, N.~Manojlovi\'c, Z.~Nagy, J.~Nunes~da Costa, and
  H.~Samtleben, ``Infinite dimensional algebras and their applications to
  quantum integrable systems,''
  \href{http://dx.doi.org/10.1088/1751-8121/41/19/190301}{{\em J. Phys. A}
  {\bfseries 41} no.~19, (2008) 190301, 2}.
  \url{http://dx.doi.org/10.1088/1751-8121/41/19/190301}. MR2452174

\bibitem{MR2470511}
P.~Kulish and V.~Lyakhovsky, ``String functions for affine {L}ie algebras
  integrable modules,'' \href{http://dx.doi.org/10.3842/SIGMA.2008.085}{{\em
  SIGMA Symmetry Integrability Geom. Methods Appl.} {\bfseries 4} (2008) Paper
  085, 18}. \url{http://dx.doi.org/10.3842/SIGMA.2008.085}. MR2470511

\bibitem{MR2759743}
E.~S. Gut$\cdot$shabash and P.~P. Kulish, ``Discrete symmetries, the {D}arboux
  transformation, and exact solutions in the {W}ess-{Z}umino-{N}ovikov-{W}itten
  model,'' \href{http://dx.doi.org/10.1007/s10958-009-9420-4}{{\em Zap. Nauchn.
  Sem. S.-Peterburg. Otdel. Mat. Inst. Steklov. (POMI)} {\bfseries 360}
  no.~Teoriya Predstavleni\u\i , Dinamicheskie Sitemy, Kombinatornye Metody.
  XVI, (2008) 139--152, 297}.
  \url{http://dx.doi.org/10.1007/s10958-009-9420-4}. MR2759743

\bibitem{MR2549451}
M.~Il'in, P.~Kulish, and V.~Lyakhovsky, ``On the properties of branching
  coefficients for affine {L}ie algebras,''
  \href{http://dx.doi.org/10.1090/S1061-0022-10-01090-3}{{\em Algebra i Analiz}
  {\bfseries 21} no.~2, (2009) 52--70}.
  \url{http://dx.doi.org/10.1090/S1061-0022-10-01090-3}. MR2549451

\bibitem{MR2722298}
P.~Aschieri, M.~Dimitrijevi\'c, P.~Kulish, F.~Lizzi, and J.~Wess,
  \href{http://dx.doi.org/10.1007/978-3-540-89793-4}{{\em Noncommutative
  spacetimes}}, vol.~774 of {\em Lecture Notes in Physics}.
\newblock Springer-Verlag, Berlin, 2009.
\newblock \url{http://dx.doi.org/10.1007/978-3-540-89793-4}.
\newblock Symmetries in noncommutative geometry and field theory. MR2722298

\bibitem{MR2662503}
P.~P. Kulish, N.~Manojlovi\'c, and Z.~Nagy, ``Symmetries of spin systems and
  {B}irman-{W}enzl-{M}urakami algebra,''
  \href{http://dx.doi.org/10.1063/1.3366259}{{\em J. Math. Phys.} {\bfseries
  51} no.~4, (2010) 043516, 15}. 
  \url{http://dx.doi.org/10.1063/1.3366259}. MR2662503

\bibitem{MR2749810}
P.~N. Bibikov and P.~P. Kulish, ``The three-magnon problem and the
  integrability of rung-dimerized spin ladders,''
  \href{http://dx.doi.org/10.1007/s10958-010-0026-7}{{\em Zap. Nauchn. Sem.
  S.-Peterburg. Otdel. Mat. Inst. Steklov. (POMI)} {\bfseries 374} no.~Voprosy
  Kvantovo\u\i \ Teorii Polya i Statistichesko\u\i \ Fiziki. 21, (2010) 44--57,
  268--269}. \url{http://dx.doi.org/10.1007/s10958-010-0026-7}. MR2749810

\bibitem{MR2749818}
M.~Ilyin, P.~Kulish, and V.~Lyakhovsky, ``Folded fans and string functions,''
  \href{http://dx.doi.org/10.1007/s10958-010-0034-7}{{\em Zap. Nauchn. Sem.
  S.-Peterburg. Otdel. Mat. Inst. Steklov. (POMI)} {\bfseries 374} no.~Voprosy
  Kvantovo\u\i \ Teorii Polya i Statistichesko\u\i \ Fiziki. 21, (2010)
  197--212, 271}. \url{http://dx.doi.org/10.1007/s10958-010-0034-7}. MR2749818

\bibitem{MR2752975}
P.~Kulish, N.~Manojlovi\'c, M.~Samsonov, and A.~Stolin, ``Bethe ansatz for the
  deformed {G}audin model,''
  \href{http://dx.doi.org/10.3176/proc.2010.4.11}{{\em Proc. Est. Acad. Sci.}
  {\bfseries 59} no.~4, (2010) 326--331}.
  \url{http://dx.doi.org/10.3176/proc.2010.4.11}. MR2752975

\bibitem{Kulish2010}
P.~P. Kulish, N.~Manojlovi{\'{c}}, and Z.~Nagy, ``Jordanian deformation of the
  open xxx spin chain,''
  \href{http://dx.doi.org/10.1007/s11232-010-0047-x}{{\em Theoretical and
  Mathematical Physics} {\bfseries 163} no.~2, (May, 2010) 644--652}.
  \url{https://doi.org/10.1007/s11232-010-0047-x}.

\bibitem{MR2775125}
P.~Kulish and A.~Mudrov, ``Twisting adjoint module algebras,''
  \href{http://dx.doi.org/10.1007/s11005-010-0454-9}{{\em Lett. Math. Phys.}
  {\bfseries 95} no.~3, (2011) 233--247}.
  \url{http://dx.doi.org/10.1007/s11005-010-0454-9}. MR2775125

\bibitem{MR3171183}
J.~Avan, P.~P. Kulish, and G.~Rollet, ``Reflection {$K$}-matrices related to
  {T}emperley-{L}ieb {$R$}-matrices,''
  \href{http://dx.doi.org/10.1007/s11232-011-0130-y}{{\em Theoret. and Math.
  Phys.} {\bfseries 169} no.~2, (2011) 1530--1538}.
  \url{http://dx.doi.org/10.1007/s11232-011-0130-y}. Russion version appears in
  Teoret. Mat. Fiz. {{\bf{1}}69} (2011), no. 2, 194--203. MR3171183

\bibitem{MR2944987}
N.~M. Bogolyubov and P.~P. Kulish, ``Exactly solvable models of nonlinear
  quantum optics,'' {\em Zap. Nauchn. Sem. S.-Peterburg. Otdel. Mat. Inst.
  Steklov. (POMI)} {\bfseries 398} no.~Voprosy Kvantovo\u\i \ Teorii Polya i
  Statistichesko\u\i \ Fiziki. 22, (2012) 26--54, 223. MR2944987

\bibitem{MR3168711}
P.~P. Kulish, V.~D. Lyakhovsky, and O.~V. Postnova, ``Multiplicity function for
  tensor powers of modules of the {$A_n$} algebra,''
  \href{http://dx.doi.org/10.1007/s11232-012-0063-0}{{\em Theoret. and Math.
  Phys.} {\bfseries 171} no.~2, (2012) 666--674}.
  \url{http://dx.doi.org/10.1007/s11232-012-0063-0}. Translation of Teoret.
  Mat. Fiz. {{\bf{1}}71} (2012), no. 2, 283--293. MR3168711

\bibitem{more29}
P.~P. Kulish, V.~D. Lyakhovsky, and O.~V. Postnova, ``{Tensor power
  decomposition. $B_n$ case},'' {\em J. Physics: Conf. Ser.} {\bfseries 343}
  no.~1, (2012) 012095.

\bibitem{MR3221506}
P.~P. Kulish, \href{http://dx.doi.org/10.1142/9789814518550_0067}{``Integrable
  spin chains and representation theory,''} in {\em Symmetries and groups in
  contemporary physics}, vol.~11 of {\em Nankai Ser. Pure Appl. Math. Theoret.
  Phys.}, pp.~487--492.
\newblock World Sci. Publ., Hackensack, NJ, 2013.
\newblock \url{http://dx.doi.org/10.1142/9789814518550_0067}. MR3221506

\bibitem{MR3301495}
J.~Avan, P.~P. Kulish, and G.~Rollet, ``Reflection matrices from
  {H}adamard-type {T}emperley-{L}ieb {$R$}-matrices,''
  \href{http://dx.doi.org/10.1007/s11232-014-0150-5}{{\em Theoret. and Math.
  Phys.} {\bfseries 179} no.~1, (2014) 387--394}.
  \url{http://dx.doi.org/10.1007/s11232-014-0150-5}. Russian version appears in
  Teoret. Mat. Fiz. {{\bf{1}}79} (2014), no. 1, 3--12. MR3301495

\bibitem{ISI:000301174100069}
P.~P. Kulish, V.~D. Lyakhovsky, and O.~V. Postnova,
  \href{http://dx.doi.org/{10.1088/1742-6596/343/1/012070}}{``{Multiplicity
  functions for tensor powers. A(n)-case},''} in {\em 7TH International
  Conference on Quantum Theory and Symmetries (QTS7)}, vol.~{343} of {\em
  {Journal of Physics Conference Series}}, {Czech Tech Univ, Fac Nucl Sci \&
  Phys Engn, Dept Math \& Phys; Bogoliubov Lab Theoret Phys Joint Inst Nucl
  Res; Acad Sci, Inst Phys}.
\newblock {2012}.
\newblock {7th International Conference on Quantum Theory and Symmetries (QTS),
  Prague, Czech Republic, Aug 07-13, 2011}. ISI:000301174100069

\bibitem{ISI:000305176100012}
P.~P. Kulish, V.~D. Lyakhovsky, and O.~V. Postnova,
  \href{http://dx.doi.org/{10.1088/1742-6596/346/1/012012}}{``{Tensor powers
  for non-simply laced Lie algebras B-2-case},''} in {\em Algebra, Geometry,
  and Mathematical Physics 2010}, {Abramov, V and Fuchs, J and Paal, E and
  Shestopalov, Y and Silvestrov, S and Stolin, A}, ed., vol.~{346} of {\em
  {Journal of Physics Conference Series}}.
\newblock {2012}.
\newblock {6th Baltic-Nordic Workshop on Algebra, Geometry, Mathematical
  Physics (AGMP), Sven Loven Ctr Marine Sci, Sweden, Oct 25-31, 
  2010}. ISI:000305176100012

\bibitem{ISI:000350984400006}
I.~Y. Aref'eva, V.~M. Buchstaber, E.~P. Velikhov, A.~B. Zhizhchenko, V.~E.
  Zakharov, I.~A. Ibragimov, S.~V. Kislyakov, V.~V. Kozlov, P.~P. Kulish, L.~N.
  Lipatov, V.~P. Maslov, V.~A. Matveev, S.~P. Novikov, Y.~S. Osipov, A.~M.
  Polyakov, V.~A. Rubakov, M.~A. Semenov-Tian-Shansky, Y.~A. Simonov, Y.~G.
  Sinai, A.~A. Slavnov, I.~A. Sokolov, L.~A. Takhtajan, V.~E. Fortov, and S.~L.
  Shatashvili, ``{Ludvig Dmitrievich Faddeev (on his 80th birthday)},''
  \href{http://dx.doi.org/{10.1070/RM2014v069n06ABEH004932}}{{\em {Russian
  Mathematical Surveys}} {\bfseries {69}} no.~{6}, ({2014}) 
  {1133--1142}}. ISI:000350984400006

\bibitem{MR3301515}
J.~Avan, T.~Fonseca, L.~Frappat, P.~P. Kulish, E.~Ragoucy, and G.~Rollet,
  ``Temperley-{L}ieb {$R$}-matrices from generalized {H}adamard matrices,''
  \href{http://dx.doi.org/10.1007/s11232-014-0138-1}{{\em Theoret. and Math.
  Phys.} {\bfseries 178} no.~2, (2014) 223--238}.
  \url{http://dx.doi.org/10.1007/s11232-014-0138-1}. Russian version appears in
  Teoret. Mat. Fiz. {{\bf{1}}78} (2014), no. 2, 255--273. MR3301515

\bibitem{MR3369904}
E.~V. Damaskinsky, P.~P. Kulish, and M.~A. Sokolov, ``On calculation of
  generating functions of {C}hebyshev polynomials in several variables,''
  \href{http://dx.doi.org/10.1063/1.4922997}{{\em J. Math. Phys.} {\bfseries
  56} no.~6, (2015) 063507, 11}. 
  \url{http://dx.doi.org/10.1063/1.4922997}. MR3369904

\bibitem{MR3589027}
P.~A. Valinevich, S.~E. Derkachev, P.~P. Kulish, and E.~M. Uvarov,
  ``Construction of eigenfunctions for a system of quantum minors of the
  monodromy matrix for an $SL(n,\Bbb{C})$-invariant spin chain,''
  \href{http://dx.doi.org/10.4213/tmf9106}{{\em Teoret. Mat. Fiz.} {\bfseries
  189} no.~2, (2016) 149--175}. \url{http://dx.doi.org/10.4213/tmf9106}. MR3589027

\end{thebibliography}
\end{document}